# Influence of chemical composition on the room temperature plasticity of *C*15 Ca-Al-Mg Laves phases

Martina Freund[1,*], Zhuocheng Xie[1,*], Pei-Ling Sun[1], Lukas Berners[1], Joshua Spille[2], Hexin Wang[1], Carsten Thomas[3], Michael Feuerbacher[3], Marta Lipinska-Chwalek[2,3], Joachim Mayer[2,3], Sandra Korte-Kerzel[1]

[1]Institute of Physical Metallurgy and Materials Physics, RWTH Aachen University

[2]Central Facility for Electron Microscopy, RWTH Aachen University

[3] Ernst Ruska-Centre for Microscopy and Spectroscopy with Electrons, Forschungszentrum Jülich GmbH

*Corresponding authors: Martina Freund, freund@imm.rwth-aachen.de; Zhuocheng Xie, xie@imm.rwth-aachen.de

# Abstract

The influence of chemical composition changes on the room temperature mechanical properties in the *C*15 $CaAl_2$ Laves phase were investigated in two off-stoichiometric compositions with 5.7 at.-% Mg addition ($Ca_{33}Al_{61}Mg_6$) and 10.8 at.-% Mg and 3.0 at.-% Ca addition ($Ca_{36}Al_{53}Mg_{11}$) and compared to the stoichiometric ($Ca_{33}Al_{67}$) composition. Cubic Ca-Al-Mg Laves phases with multiple crystallographic orientations were characterised and deformed using nanoindentation. The hardness and indentation modulus were measured to be 4.1 ± 0.3 GPa and 71.3 ± 1.5 GPa for $Ca_{36}Al_{53}Mg_{11}$, 4.6 ± 0.2 GPa and 80.4 ± 3.8 GPa for $Ca_{33}Al_{61}Mg_6$ and 4.9 ± 0.3 GPa and 85.5 ± 4.0 GPa for $Ca_{33}Al_{67}$, respectively. The resulting surface traces as well as slip and crack planes, were distinguished on the indentation surfaces, revealing the activation of several different {11n} slip systems, as further confirmed by conventional transmission electron microscopic observations. Additionally, the deformation mechanisms and corresponding energy barriers of activated slip systems were evaluated by atomistic simulations.

# 1. Introduction

Laves phases present as some of the most frequent intermetallic structures and are attractive for strengthening of structural materials. They consist of two atoms with different atomic radii, with an ideal radius ratio in AB$_2$ binary Laves phases of $r_A/r_B = 1.225$ [12-17].

Due to the TCP structure, the introduction of macroscopic plasticity is hindered by the inherent resistance to dislocation motion and hence there exists a lack of knowledge regarding the plastic behaviour and rare documentation of dislocation motion below the brittle to ductile transition temperature (BDTT), usually encountered at around 0.6·T$_H$ (where T$_H$ is the homologous temperature) for Laves phases [19-23].

For the cubic C15 Laves phases, a review of the commonly reported slip systems and the corresponding BDTT temperatures was included in a previous publication [24]. Dislocation slip is predominantly reported on {111} planes at and above the BDTT [22, 23, 25-33]. Several studies using micropillar compression to introduce plasticity at room temperature also revealed slip on {111} planes [34-38]. However, investigations on CaAl$_2$ Laves phases have shown significant contributions from {112} slip planes to plasticity [24].

Theoretical studies using atomistic simulations have shed further light on plasticity mechanisms in Laves phases [39-41]. Dislocation motion by synchro-shear [16, 42, 43] was demonstrated as the most energetically favourable slip mechanism on the basal plane in the hexagonal *C*14 CaMg$_2$ Laves phase [39], while kink-pair nucleation and propagation are preferred in both *C*14 CaMg$_2$ and *C*15 CaAl$_2$ Laves phases [41] and thermal assistance was found to be indispensable in activating the motion of synchro-Shockley dislocations, implying that this type of plastic event is impeded at low temperatures [40]. Point defects, such as vacancies, introduced with deviations from the stoichiometric AB$_2$ composition [18], significantly affect dislocation motion energy barriers [13].

To better understand the plasticity of Laves phases with off-stoichiometric compositions, we investigate the deformation of the *C*15 CaAl$_2$ Laves phase with different Mg contents (presumably replacing Al in the lattice) and a change in Ca concentration. This study aims to elucidate the impact of compositional changes on plasticity in Laves phases, building on previous research on compositional stability and the effects of chemical composition on plasticity.

## 1.1 Influence of chemistry on Laves phase stability and plasticity

The knowledge of the influence of chemistry on the mechanical behaviour and phase stability of Laves phases, particularly those with well-defined homogeneity ranges, can be helpful for a specific property manipulation. Amerioun et al. [44] investigated the structural changes in

Laves phases when substituting Al atoms with Mg. They found transformations from the $C$15 Laves phase (homogeneity range from $CaAl_2$ to $CaAl_{1.76}Mg_{0.24}$) to the $C$36 phase (homogeneity range from $CaAl_{1.34}Mg_{0.66}$ to $CaAl_{0.93}Mg_{1.07}$), and further to the hexagonal $C$14 Laves phase (homogeneity range from $CaAl_{0.49}Mg_{1.51}$ to $CaMg_2$).

The mechanical behaviour of Laves phases is significantly influenced by composition, with deviations from stoichiometry leading to the insertion of various types of defects into the lattice, such as anti-site atoms and vacancies [14, 48, 49, 51, 57]. Altering the composition away from the ideal binary stoichiometry, either by varying the internal chemistry or adding alloying elements, results in different defect structures [14, 15, 45, 46, 48, 49, 51, 58-60]. This can lead to a distortion of the ideal lattice and changes in lattice parameters, subsequently affecting mechanical properties [18]. The introduction of a ternary element also plays a crucial role, with the atomic radius and quantity of the element determining how it is incorporated into the lattice [13, 14, 18]. Moreover, the radius ratio in cubic Laves phases significantly affects dislocation motion on {111} planes, influencing the atomic free volume and dislocation mobility. The presence of anti-site atoms or vacancies can impede or facilitate dislocation motion, highlighting the intricate relationship between composition, defect structures, and mechanical behaviour in Laves phases [27, 58].

In terms of the changes in mechanical properties resulting from a changing Laves phase composition, we focus first on the effect altering the ratio of A and B atoms in the binary $AB_2$ Laves phases, before considering the inclusion of a third element:

Shields et al. [56] explored the microhardness of quenched binary rare-earth cubic $CeNi_2$ Laves phases and found that strain fields induced by vacancies in the lattice could result in softening at high temperatures and hardening after quenching at low temperatures.

Similar investigations by Chen et al. [48, 49] on cubic $HfCo_2$ Laves phases demonstrated a decrease in lattice parameter and mechanical properties with increasing Co content, attributed to Co atom substitutions and vacancies. These studies highlight the complex relationship between composition, lattice parameters, and mechanical properties in Laves phases.

Moreover, deviations from stoichiometry in hexagonal $MgZn_2$ Laves phases [12, 65-67] have been associated with decreased hardness, yield stress, and dislocation velocity, indicating the presence of various defects such as anti-site atoms and vacancies.

Zhu et al. [14] observed hardening in both over- and under-stoichiometric compositions at room temperature (RT) in cubic $NbCr_2$, $NbCo_2$, and hexagonal $NbFe_2$ alloys, attributing it to anti-site defects in the absence of constitutional vacancies. However, Luo et al. [35] found increased hardness towards the stoichiometric composition in the cubic $C$15 $NbCo_2$ Laves phase. Interestingly, micropillar compression tests by the same authors across the same compositional range of the $C15$ $NbCo_2$ phase revealed a constant critical resolved shear stress for slip on {111} planes [36].

The effect of different point defects on dislocation mobility therefore remains unclear based on experimental literature. This complexity arises because distinguishing anti-site defects and vacancies in the ordered structure of Laves phases requires elaborate experimental investigations. Consequently, making in-depth connections between point defect types, concentrations, and their interactions with dislocations necessitates a combination of many experiments. Modelling efforts have started to contribute to this field, for example by predicting the formation energies of the different defect types from density functional theory [68] and studying their interaction with dislocations in atomistic simulations [41].

Phase transformations between different Laves polytypes are also closely related to mechanical deformation, with stacking fault formation occurring due to the movement of synchro-Shockley dislocations on adjacent slip planes [16, 42, 43]. This is enabled by the construction of the three different Laves structures, which are stacking variants normal to their characteristic triple layer lying parallel to the {111} or (0001) planes in the cubic and hexagonal phases, respectively [55, 69]. Importantly, Laves phases exhibit such phase transformations between polytypes through the accumulation of stacking faults (SFs) as a result of deviations from the ideal composition [27, 60, 70].

RT compression tests of the $ZrFe_2$ Laves phase have shown phase transformations from $C36$ to $C15$ under compressive strain, driven by the motion of synchro-Shockley dislocations on {111} planes [52]. Manipulating Laves phase composition therefore not only alters plasticity based on point defects and lowers the energy barrier for dislocation motion but also simplifies phase transformations via SFs [16, 42, 43].

The influence of a third element, on the mechanical properties of Laves phases has been extensively studied, primarily focusing on phases formed by transition metals. These investigations have revealed intricate relationships between composition and mechanical behaviour.

In the Fe-Nb-Ni ternary system, Fe-rich $NbFe_2$ Laves phases experienced solid solution softening with increasing Ni content, leading to reduced hardness and elastic modulus. Similar observations were made in the Nb-Cr-Ti system, where stacking faults and dipoles were observed, affecting mechanical behaviour.

Chen et al. [57] explored the effects of ternary alloying elements (Nb, V, Mo) on cubic $TiCr_2$ Laves phase. They found that while all three elements increased hardness, V and Mo led to increased fracture toughness, whereas Nb decreased it. Additionally, V and Mo substitutions on specific sites affected lattice strain and free volume, influencing dislocation motion. Similar investigations by Takasugi et al. [58] on cubic $NbCr_2$ Laves phase showed that V and Mo substitutions resulted in either hardening or softening, depending on the specific sites they occupied. The addition of W and Mo induced phase transformations and stacking faults, affecting mechanical properties differently. Yoshida et al. [45] expanded this work to also including 5% W replacing either Nb or Cr in the alloys' composition. They found significant hardening when substituting Cr but only weak hardening when W was added instead of Nb. However, this study lacked measurements of lattice parameters to validate the hypothetical replacements based on alloy composition. The authors observed stacking faults (SFs) and initial phase transformations to hexagonal polytypes upon the addition of W and Mo, which was not observed with V addition. Consequently, there was no apparent correlation between the presence of SFs on {111} planes (seen in Mo and W-containing alloys) and overall softening (observed only in (Nb,Mo)Cr$_2$). Further research by Thoma et al. [59] investigated the effects of V addition to cubic NbCr$_2$ Laves phase, showing changes in moduli and transition temperatures.

Overall, these investigations highlight the complexity of ternary alloying effects on Laves phases, with mechanisms such as synchro-shear emerging as crucial for understanding mechanical behaviour. However, gaps in understanding persist, particularly regarding low-temperature plasticity mechanisms and their dependence on composition. Bridging experimental and computational approaches could provide deeper insights into these mechanisms, addressing fundamental questions and guiding the design of novel Laves phase materials with tailored mechanical properties.

Studies also considered equal substitution of A- and B-sites, such as Nb addition to HfV$_2$, [60, 73, 74], resulting in improved deformability attributed to the creation of free volume on {111} planes. Twinning was identified as the primary mechanism of low-temperature deformation.

In summary, while the hypothesis of creating free volume in the triple layer through ternary alloying has been explored in several studies, conclusive evidence is lacking. However, there is growing support for the role of synchro-shear in relating site-specific element substitution to mechanical properties. Other dislocation mechanisms are often overlooked in experimental studies, which primarily focus on determining the type of present point defects or identifying deformation mechanisms based on mechanical properties alone. To address these gaps, a closer integration of experimental and computational approaches is essential. By combining these methods, researchers can better understand the underlying dislocation mechanisms and identify potential defects. This integrated approach is particularly crucial given the increasing complexity of investigations involving varying compositions and exploring fundamental mechanisms at perfect stoichiometry.

From this review of the existing literature, we identify two current gaps in knowledge with respect to Laves phase plasticity, which we aim to address in this work: a general lack of understanding of low temperature plasticity mechanisms and how these are affected by changes in composition.

Advancements in nanomechanical testing now offer the capability to introduce plasticity without being constrained by the BDTT, as fracture can be successfully suppressed [37]. By utilizing a combination of scanning electron microscopy (SEM) and TEM to examine the slip planes and Burgers vectors of dislocations, it becomes possible to analyse the resulting defect structures and relate these to critical stresses for dislocation-mediated plasticity at room temperature. The impact of adding ternary alloying elements and deviations from stoichiometric compositions in Laves phases remains a debated topic. This study therefore aims to evaluate the influence of chemical composition by comparing it with the authors' previous investigation on the room temperature plasticity of the stoichiometric $C$15 $CaAl_2$ Laves phase [24]. In that study, deformation primarily occurred due to dislocation motion on {111} and {112} slip planes, which may be expected to change with the addition of Mg and/or Ca atoms. Atomistic simulations serve to rationalise experimental observations on composition-dependent mechanical properties and activation of slip systems, particularly focusing on the associated energy barriers. The present analysis aims to delve deeper into this change, seeking to bridge the gap in knowledge by understanding the influence of ternary alloying elements on deformation mechanisms in cubic Laves phases.

## 2. Experimental methods

### 2.1 Sample preparation

In this work, two off-stoichiometric compositions of the $C$15 $CaAl_2$ Laves phase were studied. They are $CaAl_2$ with 5.7 at.% Mg addition, $Ca_{33}Al_{61}Mg_6$, and 10.8 at.% Mg and 3.0 at.% Ca addition, $Ca_{36}Al_{53}Mg_{11}$.

A first ingot was prepared by flux-growth. A melt of composition $Ca_{32.7}Al_{31.9}Mg_{35.4}$ was sealed in a tantalum crucible under 500 mbar argon atmosphere, which was in turn sealed in a protecting quartz tube. This ensemble was placed in a buffered box furnace and heated to a temperature of 870°C where it was kept for 2 hours for homogenization. Cooling was carried out by lowering the temperature to 850°C at a rate of 10°C/h and subsequently to 750°C at a rate of 1°C/h. Then the ensemble was rapidly removed from the furnace, and the remaining melt and the solidified part were separated by centrifugation. Finally, it was slowly cooled down to room temperature within about 2 to 3 h.

A cross cut of the resulting ingot was prepared for phase analysis in the scanning electron microscope (SEM) (Figure 1). It contained several phases, the compositions of which were determined using energy dispersive X-ray spectroscopy (EDX) (FIB, Helios Nanolab 600i, FEI) measurements. A large volume fraction of $Ca_{33}Al_{61}Mg_6$ regions was found, the composition was determined as 61.4 ± 0.2 at.-% Al, 32.9 ± 0.2 at.-% Ca and 5.7 ± 0.1 at.-% Mg.

A second ingot was prepared by Bridgman growth. A tantalum ampoule was charged with raw elements of composition $Ca_{32.7}Al_{31.9}Mg_{35.4}$ and sealed under 600 mbar argon atmosphere. The ampoule was placed in a vertical Bridgman furnace on a water-cooled cold finger and heated to 900 °C, where it was kept 1 h for homogenization. Then the growth process was run by withdrawing the melt out of the hot zone of the furnace at a rate of 5 mm/h, which was proceeded for 40 h, i.e. 200 mm.

A cross cut of the resulting ingot again displayed the presence of several phases. Regions of $Ca_{36}Al_{53}Mg_{11}$ were found with a composition according to EDX measurements of 52.9 ± 0.7 at.% Al, 36.3 ± 0.1 at.% Ca and 10.8 ± 0.8 at.% Mg.

For metallographic preparation the samples were embedded in copper paste to provide a stable sample environment and to hinder fracture or porosity. Afterwards the sample was ground using with 1200 to 4000 grit SiC paper. An additional grinding step followed with plates of 3 µm and 1 µm and polishing in 4 steps with 6, 3, 1, and 0.25 µm diamond paste with isopropanol with 5% PEG as lubricant. A final OPA-polish (5% silica dissolved in isopropanol and added PEG) and cleaning step with water and dishwashing liquid followed.

The use of electron backscatter diffraction (EBSD) (Hikari, EDAX Inc.) at an accelerating voltage of 20 kV and a current of 5.5 nA allowed us to identify the grain orientation of the cubic $CaAl_2$ phase for both samples, and also identify, for $Ca_{33}Al_{61}Mg_6$, the pure Mg matrix lying around the individual $CaAl_2$ + Mg areas (Figure 1). This data was further analysed using OIM analysis[TM] (EDAX Inc.) to obtain the Schmid factors of the grains used for an approximation of first activated slip systems, assuming uniaxial compression along the surface normal.

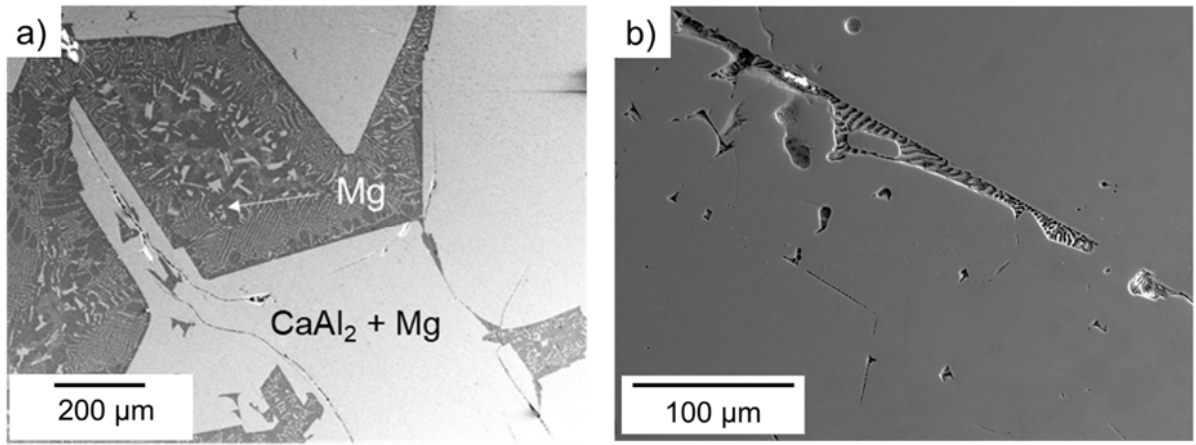

Figure 1: Microstructure of a) $Ca_{33}Al_{61}Mg_6$. The large $CaAl_2$ + Mg islands are labeled and the Mg matrix shown with the white arrow. The eutectic structures within this Mg matrix were not further analysed. In b) the microstructure of $Ca_{36}Al_{53}Mg_{11}$ the main structure was determined to be C15+ Mg, the eutectic structure was not analysed further. Both SE images were taken, using SEM in depth mode at an accelerating voltage of 5kV at a current of 200 pA.

For imaging of the plastic zone around the indents, secondary electron (SE) images were acquired by SEM in high-resolution mode (CLARA, Tescan, Brno, Czech Republic) with an accelerating voltage of 2 kV at a current of 100 pA.

## 2.2 Nanoindentation experiments

Continuous stiffness measurement (CSM) nanoindentation tests (iNano, Nanomechanics Inc., TN, USA) were performed using a diamond Berkovich tip (Synton-MDP AG, CH). Prior to testing, the diamond area function (DAF) and frame stiffness of the indenter tip were calibrated according to the Oliver and Pharr method [75, 76]. The determination of hardness and indentation modulus for all investigated areas was performed at indentation depths between 350 and 500 nm and assuming a constant Poisson's ratio of 0.3. For $Ca_{33}Al_{61}Mg_6$, eight grain orientations were indented with two arrays of 5 x 5 indents with a relative rotation around the surface normal of 30°, to investigate the influence of the tip geometry on the slip and crack behaviour. All nanoindentation tests were conducted with a constant strain rate of 0.2 1/s up to a maximum depth of 500 nm. Slip traces observed in the direct vicinity of the indents were analysed by correlating the information from EBSD data and corresponding SE-images.

## 2.3 TEM experiments

In total, three TEM lamellae were milled in two different orientations by site specific FIB milling in both off-stoichiometric materials, two for $Ca_{33}Al_{61}Mg_6$ and one for $Ca_{36}Al_{53}Mg_{11}$. All were then analysed by TEM (JEOL JEM-F200 and FEI Tecnai F20 TEM) with a double tilt holder. Under two-beam conditions, a $\vec{g} \cdot \vec{b}$ analysis, with $\vec{g}$ is the diffraction vector and $\vec{b}$ is the Burgers vector of the dislocations, was applied to determine the Burgers vectors of the dislocations. In addition, selected area electron diffraction (SAED) in combination with images was also applied to confirm the presence of the cubic $C$15 Laves phase structure underneath the indent and acquire orientation information for slip plane identification.

## 2.4 Atomistic simulations

The atomistic simulations in this work were performed using the software package LAMMPS [78]. The interatomic interactions were modelled by a machine-learning moment tensor potential (MTP) [79, 80] for Al-Ca-Mg [81]. This potential provides more accurate prediction for the lattice parameter, elastic constants and stacking fault energy of the $C$15 $CaAl_2$ Laves phase than a modified embedded atom method (MEAM) potential for Al-Ca-Mg [82] when compared with experimental data and ab-initio calculations (see Table S1 in the supplementary material). $C$15 $CaAl_2$ atomistic samples were constructed using Atomsk [83], and off-stoichiometric compositions were generated by randomly substituting atoms using five different random seeds for each composition. To model a chemical composition close to that of $Ca_{33}Al_{61}Mg_6$ in the experiments, the Ca content remains at 33 at.% and Mg solutes (3, 6 and 9 at.%) substitute the Al sublattice. For $Ca_{36}Al_{53}Mg_{11}$, the Ca sublattice remains intact and additional 3 at-% Ca and 11 at.% Mg solutes substitute the Al sublattice.

Generalized stacking fault energy (GSFE) lines and surfaces were calculated by incrementally shifting one-half of the crystal across the slip direction and plane. Periodic boundary conditions (PBC) were applied in the directions parallel to the slip plane. Fixed boundary conditions were applied along the slip plane normal, where the outermost atomic layers with a thickness of 14 Å at both top and bottom were fixed. The aspect ratio of the simulation samples is greater than 10. The crystal was relaxed in the direction perpendicular to the slip plane after each displacement step using the FIRE [84, 85] algorithm with a force tolerance of $10^{-8}$ eV/Å. Climbing image nudged elastic band (NEB) [86, 87] calculations were performed on the simulation setup with the same boundary conditions as mentioned above to find saddle points and minimum energy paths (MEPs) of slip events. The spring constants for parallel and perpendicular nudging forces are both 1.0 eV/Å$^2$. Quickmin [88] was used to minimize the energies across all replicas until the force norm was below 0.01 eV/Å.

# 3. Results

For the evaluation of the effects of chemical composition changes (as variation of Mg and/or Ca content) on the mechanical properties in the cubic CaAl$_2$ Laves phase at ambient temperature, a combination of nanomechanical testing with EDX, EBSD and TEM experiments was applied. With the employed approach, hardness and indentation modulus variations determined from nanoindentation tests were compared with local chemistry, orientation, and resulting deformation defects. In particular, EDX and TEM were used to measure the overall chemistry. The slip traces formed in the vicinity of the indents were analysed for the different grain orientations (Table 1 and Figure 2) to statistically reveal the activated slip planes. Moreover, TEM investigations were performed in selected areas, where dislocations were introduced by the nanomechanical tests, to analyse the Burgers vectors directions of these dislocations. Atomistic simulations were conducted to investigate the influence of the chemical composition on the mechanical properties and slip activation.

## 3.1 Nanoindentation

For Ca$_{33}$Al$_{61}$Mg$_6$, the average hardness of the indentation experiment is calculated to be 4.6 ± 0.2 GPa, and the average indentation modulus is 80.7 ± 3.0 GPa. The average values calculated for Ca$_{36}$Al$_{53}$Mg$_{11}$ were determined to be 4.1 œ 0.3 GPa and 71.3 œ 1.5 GPa, respectively.

Because of the large grains for the indented orientations of Ca$_{33}$Al$_{61}$Mg$_6$, we were able to extract orientation dependent hardness and indentation modulus data. Figure 2 shows all measured grain orientations located in an inverse pole figure (IPF). Table 1 gives information on the chemical compositions, hardness and indentation moduli of all analysed orientations for Ca$_{33}$Al$_{61}$Mg$_6$. The hardness values and indentation moduli are calculated as an average over both indentation experiments for each orientation (random and +30° rotated states). This was done because of their minor deviation.

Table 1: Summary of the chemistry, hardness and indentation modulus for all orientations for Ca$_{33}$Al$_{61}$Mg$_6$, whereby the orientation numbers relate to the numeration of these orientations in the IPF of Figure 2.

| | Chemical composition | | | Hardness [GPa] | Indentation Modulus [GPa] |
|---|---|---|---|---|---|
| | Mg [at.%] | Al [at.%] | Ca [at.%] | | |
| I | 5.86 ± 0.15 | 61.24 ± 0.17 | 32.90 ± 0.16 | 4.72 ± 0.35 | 78.34 ± 2.11 |
| II | 5.84 ± 0.17 | 61.48 ± 0.18 | 32.68 ± 0.14 | 4.57 ± 0.20 | 80.79 ± 1.89 |
| III | 6.15 ±0.15 | 61.07 ± 0.21 | 32.78 ± 0.18 | 5.48 ± 0.39 | 89.48 ± 1.94 |
| IV | 5.94 ± 0.23 | 61.54 ± 0.21 | 32.52 ± 0.14 | 4.77 ± 0.11 | 77.05 ± 1.48 |
| V | 5.89 ± 0.22 | 61.09 ± 0.32 | 33.02 ± 0.17 | 4.84 ± 0.18 | 81.65 ± 1.55 |
| VI | 5.79 ± 0.28 | 61.19 ± 0.29 | 33.02 ± 0.14 | 4.79 ± 0.63 | 78.94 ± 1.75 |
| VII | 5.86 ± 0.11 | 61.16 ± 0.12 | 32.98 ±0.13 | 4.83 ± 0.27 | 80.44 ± 1.52 |
| VIII | 6.02 ± 0.14 | 60.89 ± 0.19 | 33.09 ± 0.16 | 4.61 ± 0.29 | 78.57 ± 1.94 |

## 3.2 Surface trace analysis

For the statistical analysis of the resulting slip and crack planes, a total of 7172 surface traces was carefully analysed following the method by Gibson et al. [77] using aligned EBSD data and SE images. The tolerance angle between the alignment of the identified slip traces and the ideal orientation of potential slip plane traces was chosen to be 3° to account for small deviations from perfect alignment during normal imaging and EBSD analysis under 70°. The plastic zones around the indentation marks showed three different morphologies: straight lines, edges and curves, the same as in the earlier publications of Laves phases and µ-phases [24, 89, 90]. We also encountered areas where only cracking occurred or no visible surface traces formed.

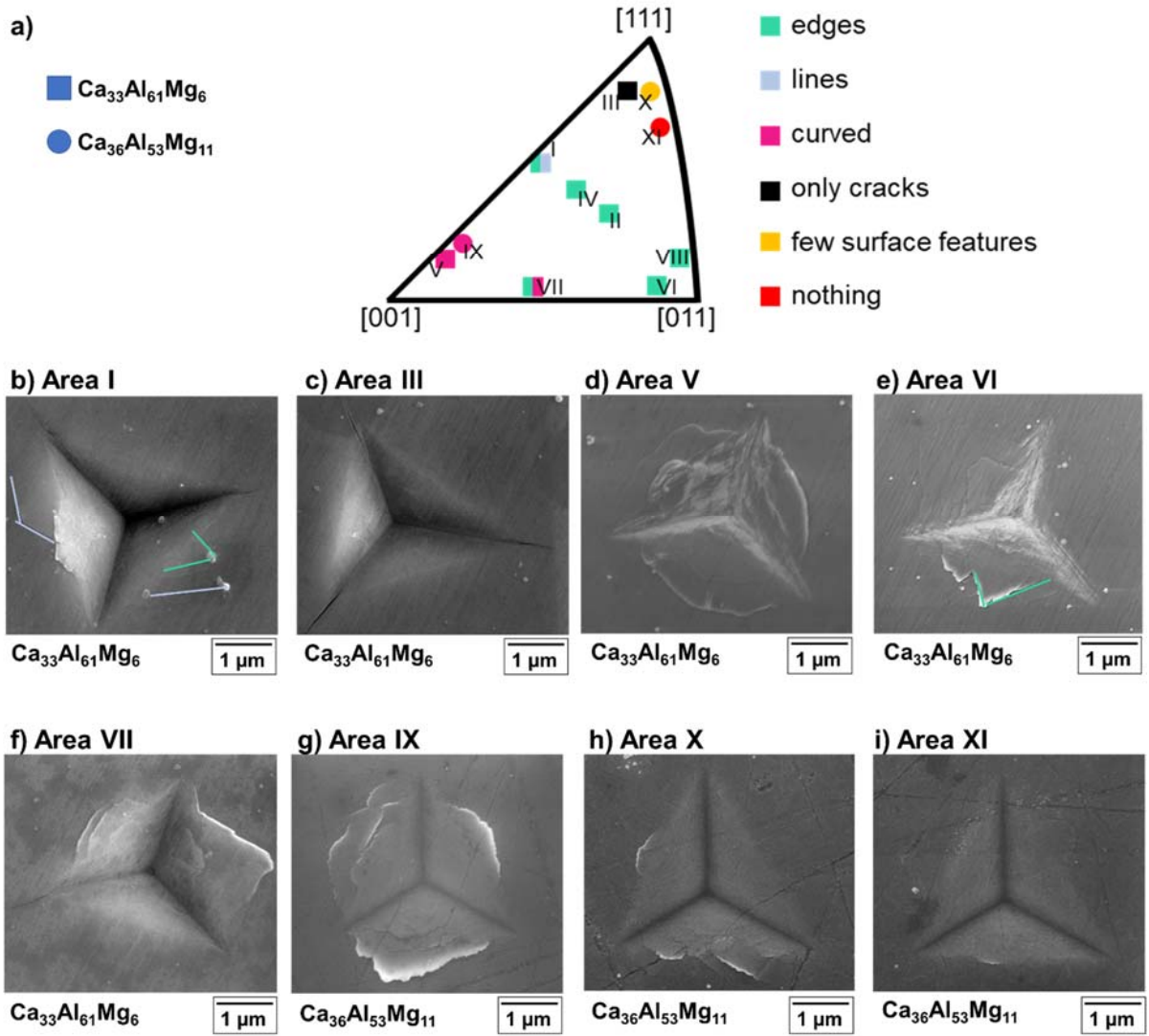

Figure 2: Different morphologies of the plastic zones around the indents, where the full square represents the resulting morphology for $Ca_{33}Al_{61}Mg_6$, and those without filling represent the morphology for $Ca_{36}Al_{53}Mg_{11}$. In a) the IPF map, the pink squares/circles indicate the orientation where curved slip lines were observed in d) for area V ($Ca_{33}Al_{61}Mg_6$) and g) for area IX ($Ca_{36}Al_{53}Mg_{11}$). Black squares show the orientation where only cracks occurred, comparable with the SE images of area III ($Ca_{33}Al_{61}Mg_6$) in c). Green squares mark the orientations where edge slip traces formed, shown for area VII and VI ($Ca_{33}Al_{61}Mg_6$) in e), f). Light blue squares highlight the orientations where straight slip lines are visible, which only occurred together with edge slip lines for area I ($Ca_{33}Al_{61}Mg_6$) displaced in b) the light blue lines show the line structure and the green lines the edge structure. Orange circles for the case that just a few surface features were observed, like in area X ($Ca_{36}Al_{53}Mg_{11}$) in h) and at least red circles where nothing happened, shown in i) for area XI ($Ca_{36}Al_{53}Mg_{11}$).

Figure 2 displays the indentation surfaces with the six distinct indentation marks and their orientations in the IPF map (see Figure 2a). Figure 2b–f shows representative morphologies of $Ca_{33}Al_{61}Mg_6$. Figure 2b shows straight lines mixed with edges around the indent in area I. Figure 2c presents only cracks appearing around the indent in area III, located near {111}. Figure 2d reveals a curved slip trace morphology in area V, oriented most closely towards {001}. Figure 2e shows edges around the indent in area VI and Figure 2f the mixture of edges and curves for area VII. The rotation of the sample does not affect any orientation.

Figure 2g–i show representative morphologies of $Ca_{36}Al_{53}Mg_{11}$. In Figure 2g, only the curved slip morphology is evident around the indent in area IX, while Figure 2i illustrates the absence of visible surface deformation in area XI. Figure 2h displays all defined morphology types observed in area X.

The analysis of the surface traces can only be performed on those traces which contain straight components, such as the straight lines (Figure 2b), cracks (Figure 2c) and parts of the edges (Figure 2e). For the curved traces, an evaluation of possible slip planes is not feasible, therefore, the resulting plasticity of area V as well as area IX and area XI (because of the absence of slip lines) (Figure 2d, g, i) is not taken into account for the statistical evaluations of the activated slip planes.

For the statistical analysis of the resulting slip traces forming around the indents, we considered in the present work more planes than in a previous study [24]: the low index {100}, {110} slip planes as well as the {111} and {11n} planes with n = 2, …, 6 and 11. This decision was based on new observations of plastic events on these slip planes in the TEM analysis in the present work. The alignment of the considered planes with respect to the unit cell is shown in Figure 3a.

Surface traces related to plasticity were analysed statistically on samples of $Ca_{33}Al_{67}$, $Ca_{33}Al_{61}Mg_6$, and $Ca_{36}Al_{53}Mg_{11}$, yielding 892, 3902, and 992 traces, respectively. Considering all possible slip systems within the defined error of the deviation angle, the total number of slip planes was approximately three times the identified surface traces, indicating an activation frequency of around 300%. This suggests multiple potential solutions per slip line due to the tolerance angle of the slip trace analysis algorithm and orientations where unique solutions may not exist. The activation frequency of slip planes remained consistent across different indenter rotations, although the number of detected surface traces varied, with more traces generally observed around 30° clockwise rotated indentations.

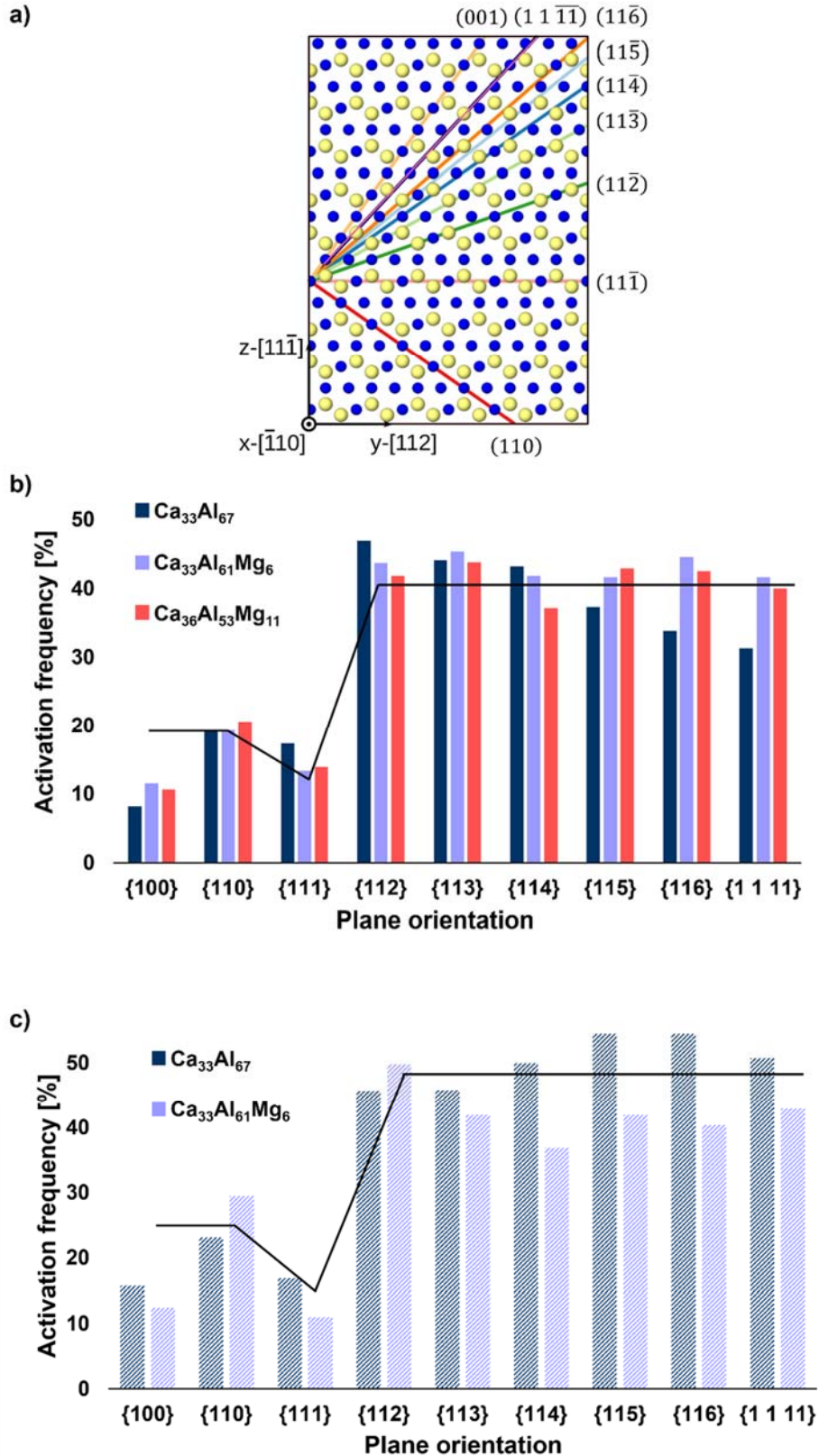

Figure 3: a) The alignment of the slip planes for the activation frequency are displaced in the cubic $C15$ $CaAl_2$ Laves phase. Additionally, the overall relative activation frequency of $Ca_{33}Al_{67}$ (in blue), $Ca_{33}Al_{61}Mg_6$ in purple and $Ca_{36}Al_{53}Mg_{11}$ in red is represented in b). The surface traces were done by analysing straight parts in respect to the {100}, {11n} planes with n={0, …, 6, 11} and the reanalysis of the stoichiometric sample from [24] (for differences in the analyses here and in [24] see Discussion). c) Total activation frequency of crack planes for the reanalysed Ca33Al67 sample (blue stripes) [24] compared to sample Ca33Al61Mg6 (purple stripes). The black line in b) and c) gives the possible slip planes per plane family.

To analyse the influence of the chemical change and the distinction of the stoichiometric Laves phase, the results of the slip line analysis of the stoichiometric $CaAl_2$ Laves phase [24] were reanalysed and plotted next to the results of the off-stoichiometric samples (Figure 3). where the total amount of activated slip systems is represented by blue bars for $Ca_{33}Al_{67}$, and by purple and light red bars for $Ca_{33}Al_{61}Mg_6$ and $Ca_{36}Al_{53}Mg_{11}$, respectively. Overall, all samples have in common, that the major slip plane activation frequency were detected on the new {11n} planes. The {110} and {100} were detected less often, irrespective of composition of the Laves phases.

Differences between the compositions emerge for the activation of the {111} plane and the higher order {11n} planes. In case of the {111} plane, activation decreases for both samples containing Mg (from 18.7 % in $Ca_{33}Al_{67}$ to 13.4% in $Ca_{33}Al_{61}Mg_6$ and 13.9% in $Ca_{36}Al_{53}Mg_{11}$). Furthermore, while the activation of the {11n} planes hovers at or above 40% (with the sole exception of {114} in $Ca_{36}Al_{53}Mg_{11}$) in the Ca-Al-Mg alloys.

Regarding the crack planes, the same procedure as for the slip planes was used. The overall activation frequency for the resulting orientations is shown in Figure 3c. As the indented areas around the indents of $Ca_{36}Al_{53}Mg_{11}$ were crack free, this sample is not included in Figure 3c. The dominant crack planes in terms of activation frequency were the {11n} planes. Overall, $Ca_{33}Al_{67}$ 394 showed cracks for 144 indents, while $Ca_{33}Al_{61}Mg_6$ revealed 992 cracks for 400 indents, resulting in a similar propensity for crack formation of 2.7 and 2.5 cracks per indent, respectively. Again, the indexation resulted in 3 times as many potential crack planes as indexable cracks in the micrographs, given a total activation frequency of the order of 300% in both cases.

Both compositions show similar distribution of crack planes with the {11n} planes at the higher activation frequencies and, as seen in the activation of slip, crack propagation on the {100}, {110} and {111} planes were also detected, but comparable to the {11n} planes at a significantly reduced frequency. For a better visualisation of the origin of multiple assignments of slip planes and the difficulty in distinguishing {11n} planes, in Figure 4 an example for $Ca_{33}Al_{61}Mg_6$ in IV of one surface trace is given, where, in addition to the {113}-{115} planes, the (011) plane could be found. The (011) plane exhibits a deviation angle of 2.18° (from the ideal trace assuming perfectly accurate determination of the surface plane and crystal orientation), a deviation angle of ($\bar{1}14$) was calculated to be 2.91°, whereas the {113} and {115} can be assigned twice with deviation angles of 1.66° and 1.74° as well as 1.56° and 1.79° respectively.

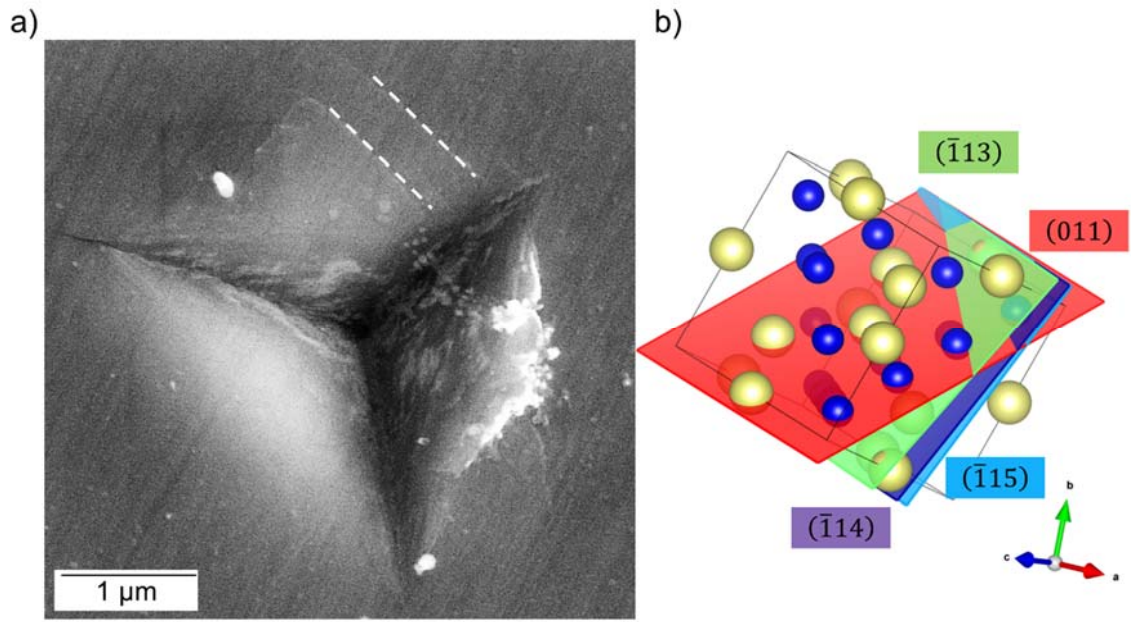

Figure 4: Exemplary visualisation of the surface analysis, showing in a) the SE image taken from orientation IV with the analysed surface trace with edge character marked with the white dashed line. In b) the CaAl2 unit cell shows the possible solution taken from the slip line detection results, but only plotting one for every plane family, with red the alignment of the (011) plane, with light green the $(\bar{1}13)$, with blue $(\bar{1}14)$ and with light blue $(\bar{1}15)$ the is drawn in the unit cell, taking VESTA for easier observation [91].

## 3.3 TEM analysis

Detailed investigation of the deformation microstructure in both non-stoichiometric "C15+Mg" Laves phases was performed by TEM analysis. Electron transparent FIB Lamellae were prepared from the material volumes located next to or directly beneath of the selected indents presented in Figure 2.

A TEM lamella was milled from orientation V in $Ca_{33}Al_{61}Mg_6$ to identify the curved slip planes (cf. Figure 2d), which cannot be identified by the surface trace analysis. Here, the TEM lamella was milled to obtain the foil plane possibly parallel to one of the {110} planes. In this way, six zone axes were accessible for imaging in TEM: the [101], [1$\bar{1}$1], [323], [312], [2$\bar{1}$3] and [3$\bar{1}$2] zone axes.

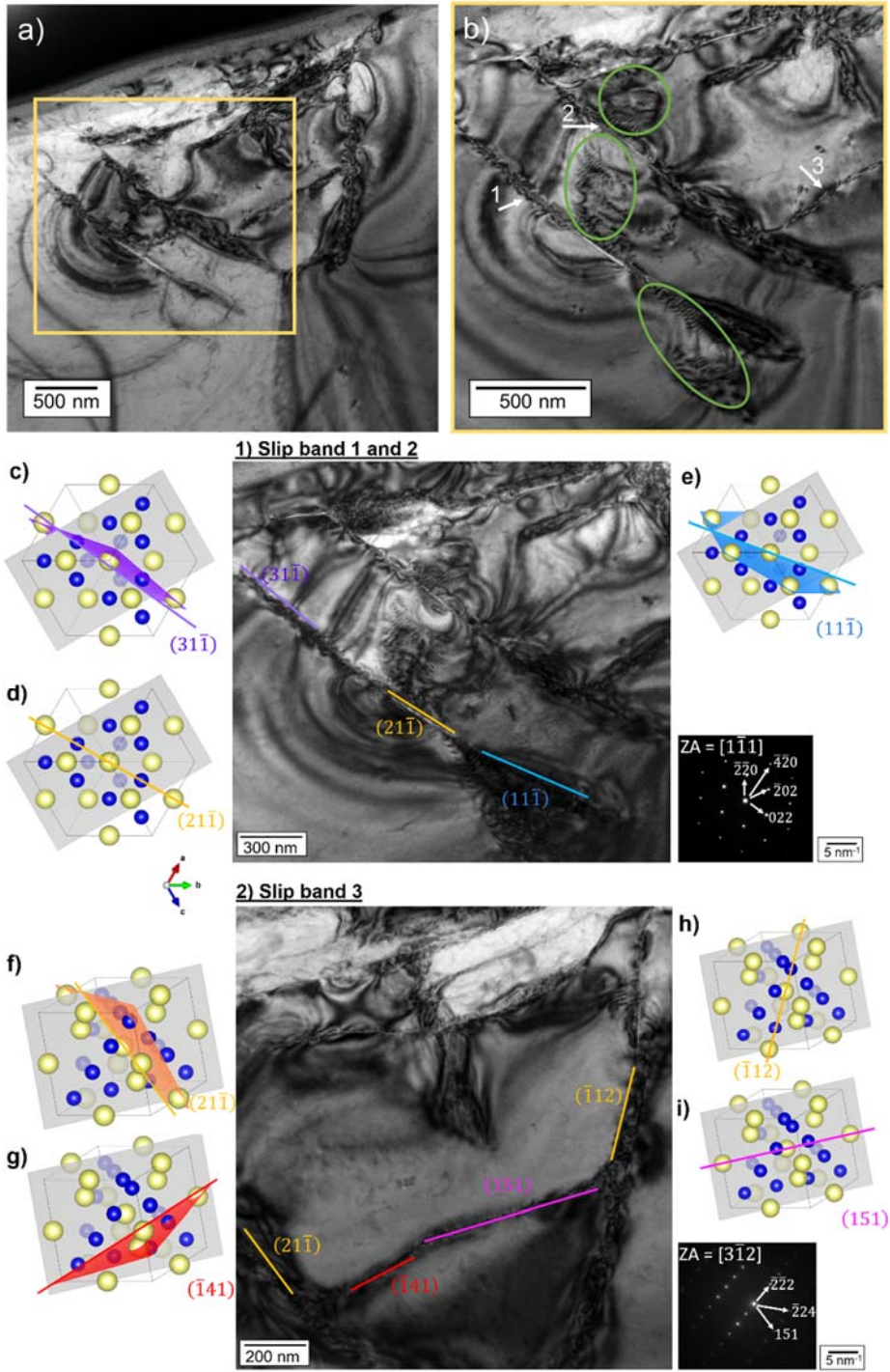

Figure 5 Dislocation structure beneath the indent zone in $Ca_{33}Al_{61}Mg_6$ from area V. a) Overview TEM BF image taken at $[1\bar{1}1]$ zone axis. The yellow square highlights the region of interest and is displaced in b) at a higher magnification. Moreover, the numbers 1 to 3 show different slip bands, which were further analysed. Activation of {11n} slip bands beneath an indentation in $Ca_{33}Al_{61}Mg_6$ in orientation V. TEM BF images taken in the 1) $[1\bar{1}1]$ zone axis and 2) $[3\bar{1}2]$ zone axis. Slip band 1 and 2 are close to edge-on at $[1\bar{1}1]$ zone axis. Slip band 3 is close to edge-on at $[3\bar{1}2]$ zone axis. c) – e) show the unit cell, lamella plane (grey plane) and the indexed slip plane (colour plane) using VESTA [91]. The colour lines are the intersections of the lamella plane and the slip plane and they are shown next to the slip bands for guidance. In 1), the slip band is observed to have segments with various orientations: the upper (purple), middle (orange) and lower (blue) sections, which are indexed to be on the $(31\bar{1})$, $(21\bar{1})$ and $(11\bar{1})$ planes, respectively. Slip band 3 in 2) consists of two segments parallel to the $(\bar{1}41)$ plane (red, f)) and the (151) plane (pink, i)). The orange lines are determined to be lie parallel to the {112} planes on $(21\bar{1})$ and $(\bar{1}12)$ planes

The image taken at the $[1\bar{1}1]$ zone axis is shown in Figure 5a. The dislocations located just under the indent are found to be confined on specific planes. A region of interest is highlighted by a yellow square in Figure 5a and presented in Figure 5b at a higher magnification. Similar dislocation dipole structures as the ones observed in the first lamella are noted here as well, which are marked by green ellipses and a circle in Figure 5b. The three "main" slip bands are highlighted in Figure 5b by the numbers 1, 2 and 3. The habit planes for individual slip band can be determined from TEM images acquired at a zone-axis sample orientation (when electron beam is parallel to the particular zone axis), if slip plane is featuring an "edge-on" orientation (i.e. slip plane is oriented parallel to the electron beam direction and therefore observed in the TEM image as possibly thin line). Slip bands 1 and 2 are found to be close to edge-on in the $[1\bar{1}1]$ zone axis, (Figure 5-1), and show the same behaviour under the analysed two beam conditions. Slip bands 3 was found to be close to edge-on in the $[3\bar{1}2]$ zone axis orientation. In order to determine the orientations of the slip bands precisely, the lamella was tilted to different zone axes, as mentioned in the previous section. The lamella (foil) plane was determined by calculating the tilting angles between foil normal direction and different zone axes to be (75 2 69). The intersection lines of the possible slip plane and the lamella plane within the unit cell are labelled by differently coloured lines in Figure 5. These coloured lines (intersections) were compared with the slip band orientations on the TEM images taken at different zone axes. The indexed slip planes are given in Figure 5c-i next to the images taken in the $[1\bar{1}1]$ zone axis and $[3\bar{1}2]$ zone axis (Figure 5-1 and 2), respectively).

With multi-beam diffraction conditions, dislocation contrast was found to obscure the slip plane orientations. Slip band 1 consists of segments with variating habit plane orientation. The bottom segment of the slip band 1 was found to be parallel to the $(11\bar{1})$ plane (trace indicated by the blue line in Figure 5e). The middle segment of slip band 1 appears parallel to the $(21\bar{1})$ plane (orange trace in Figure 5d), while the top segment lies parallel to the $(31\bar{1})$ plane (purple line in Figure 5c).

Figure 5-2 shows slip band 3 imaged in the $[3\bar{1}2]$ zone. This slip band consists of different segments, which lie on different slip planes. The left segment lies parallel to the $(\bar{1}41)$ plane (shown in red in Figure 5g), while the right segment aligns with the (151) plane trace (shown in pink in Figure 5i). In addition to slip band 3, two other slip bands on both the left and right sides were also analysed (Figure 5f and h) and assigned to the $(21\bar{1})$ and $(\bar{1}12)$ slip planes, respectively.

The deformation structures below an indent in orientation IX in $Ca_{36}Al_{53}Mg_{11}$ were also examined by TEM (Figure 6). The lamella was milled to align its orientation close to that of the lamella from $Ca_{33}Al_{61}Mg_6$ in Figure 5 and Figure 6 where the lamella plane coincides with one of the {110} planes. The overview TEM BF image of this lamella from $Ca_{36}Al_{53}Mg_{11}$ is displaced in Figure 6a, where the dashed square shows the region of interest containing dislocation structures investigated further in detail. This region also shows at a higher magnification in Figure 6b along the [101]. The microstructure under the indent is similar to that of the first lamella of $Ca_{33}Al_{61}Mg_6$ with a homogenous plastic zone in which multiple slip systems are activated. Slightly away from the indent, deformation is again found to be confined on specific planes. Some of the slip bands are seen close to edge-on, while many of them are not edge-on. To characterise the orientations of the slip bands, the lamella was tilted to six zone axes, [101], [1$\bar{1}$2], [211], [213], [1$\bar{1}$1], and [103]. The slip bands have a higher visible density of dislocations in some zone axes in Figure 6b as well as under many two-beam conditions (shown in Figure 10). This makes it difficult to determine the slip active planes in these slip bands and indicates that there may be more than one set of dislocations on the slip bands. Therefore, a more in-depth analysis was performed under different two-beam conditions and will be shown in the following sections.

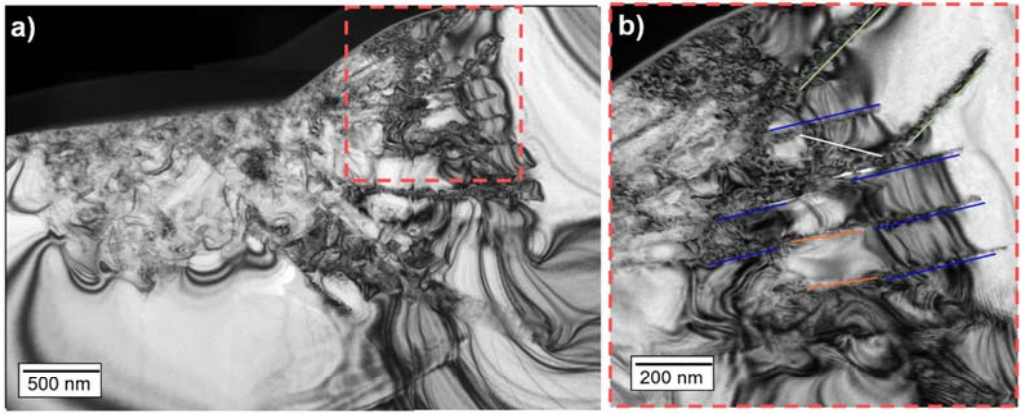
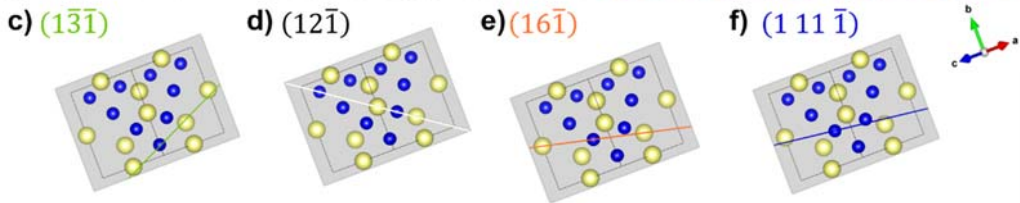
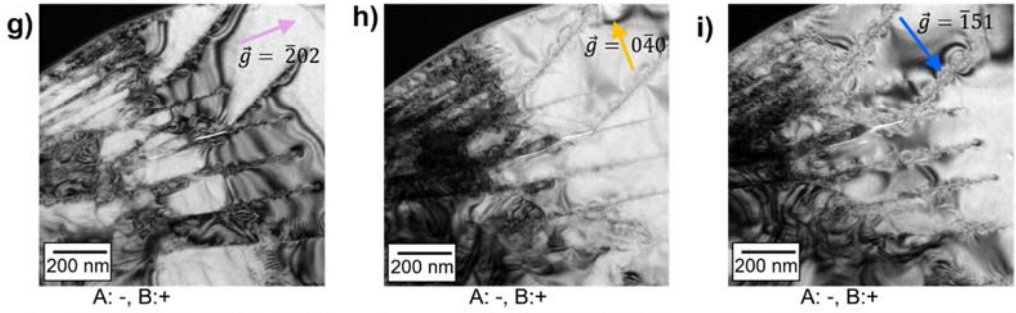
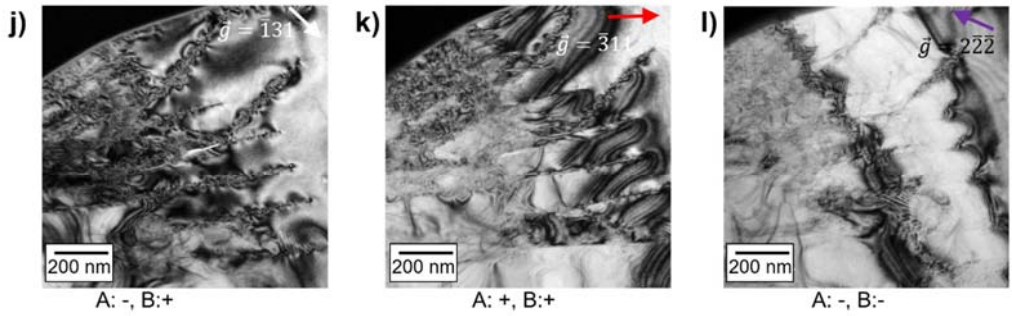
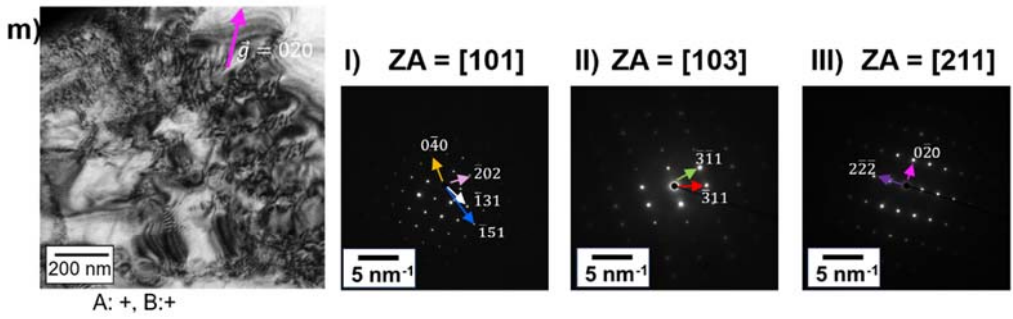
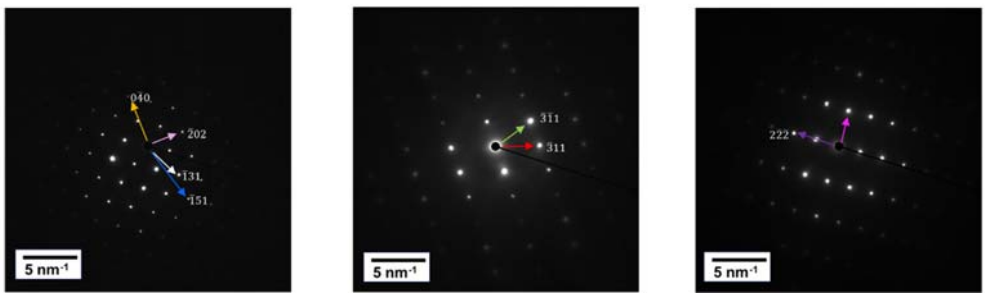

Figure 6: Deformed microstructure beneath the indent zone in $Ca_{36}Al_{53}Mg_{11}$. a) TEM BF overview image of the lamella at [101] zone axis, of area IX (Figure 2). The area marked by the dashed square is shown at b) a higher magnification. The assignment of the slip bands is additionally visualised with coloured lines, and these are shown in c)-f) orientated in the $CaAl_2$ unit cell, shown with VESTA [91]. The orange ones are displayed in e) and are on the $(16\bar{1})$ plane, the blue ones are in f) lying parallel to the $(1\ 11\ \bar{1})$ plane. In c) the light green planes are parallel to the $(1\bar{3}\bar{1})$ plane and d) goes with the white line in the BF image and the $(12\bar{1})$ plane. The applied two-beam conditions (a) – h)) and their zone axes I-III)). a)-d) belong to I) the [101] zone axis, whereby the g-vectors are shown in the in the zone axis image with the different colours and also in the BF images, the corresponding $\vec{g}$-vectors are: a) $(\bar{2}02)$ in light pink (I), b) $(0\bar{4}0)$ in yellow (I),c) $(\bar{1}51)$ in blue (I) and d) $(\bar{1}31)$ in white (I). Also, the visibility (+) or invisibility (-) of the dislocations are listed under the images. e) were taken at [103] zone axis II) using the e) $(\bar{3}11)$ $\vec{g}$-vector in green (II), f) and g) are taken in [211] zone axis III) with f) $(2\bar{2}\bar{2})$ $\vec{g}$-vector in purple (III)) and g) $(0\bar{2}0)$ $\vec{g}$-vector in pink (III)).

The slip bands imaged along the [101] zone axis and the corresponding SAD pattern are displayed in Figure 6b. At [101] zone axis, some slip bands are observed to be close to edge-on. The upper two slip bands are found to be parallel to the $(1\bar{3}\bar{1})$ plane (Figure 6c), while a $(12\bar{1})$ plane (Figure 6d) appears between the two $(1\bar{3}\bar{1})$ slip bands. In the middle of the image, several parallel slip bands can be seen. These bands again consist of segments with slightly different orientations (Figure 6e), which are indexed to be on the $(16\bar{1})$ and $(1\ 11\ \bar{1})$ planes (illustrated in Figure 6e and f, respectively). This feature of slip band is the same as observed in $Ca_{33}Al_{61}Mg_6$, which indicates that a similar deformation mechanism may have taken place in both samples.

As noted above, there is often more than one set of dislocations on the slip bands and the images taken exactly along the zone axes have a high strain contrast resulting from all of the dislocations. Therefore, two-beam conditions were applied to selectively view individual sets of dislocations based on their Burgers vector $\vec{b}$. There are two sets of dislocations, which are named "dislocation A" and "dislocation B", on a given slip plane. An analysis of $\vec{g} \cdot \vec{b}$ was conducted on two sets of dislocations. Figure 6 shows the seven specific two-beam conditions applied, namely $(\bar{2}02)$, $(0\bar{4}0)$, $(\bar{1}51)$ and $(\bar{1}31)$ $\vec{g}$-vectors in the [101] zone axis, $(\bar{3}11)$ $\vec{g}$-vector in the [103] zone axis and $(2\bar{2}\bar{2})$ and $(0\bar{2}0)$ $\vec{g}$-vectors in [211] zone axis. Dislocations A are invisible under all the two-beam conditions at the [101] zone axis and are observable under three of the applied two-beam conditions: $(\bar{3}11)$ $\vec{g}$-vectors at [103] zone axis and $(0\bar{2}0)$ $\vec{g}$-vector from the [211] zone axis. Consequently, the Burgers vector of dislocations A can be indexed to be $\frac{1}{2}[101]$. Dislocations B are visible under 6 two-beam conditions and are invisible with the $(2\bar{2}\bar{2})$ $\vec{g}$-vector from the [211] zone axis. Accordingly, the Burgers vector of dislocations B can be indexed as $\frac{1}{2}[110]$ or $\frac{1}{2}[01\bar{1}]$

At the bottom part of the lamella, some stacking faults were observed (Figure 7) which presumably did not contribute to or originate from the plastic deformation during indentation. They present as straight lines arranged in parallel and at a relative angle of 70° viewed along the

[101] zone axis, as indicated in Figure 7. The alignment of these lines is visualised in the cubic CaAl$_2$ unit cell; they correspond to the $(11\bar{1})$ and $(1\bar{1}\bar{1})$ planes.

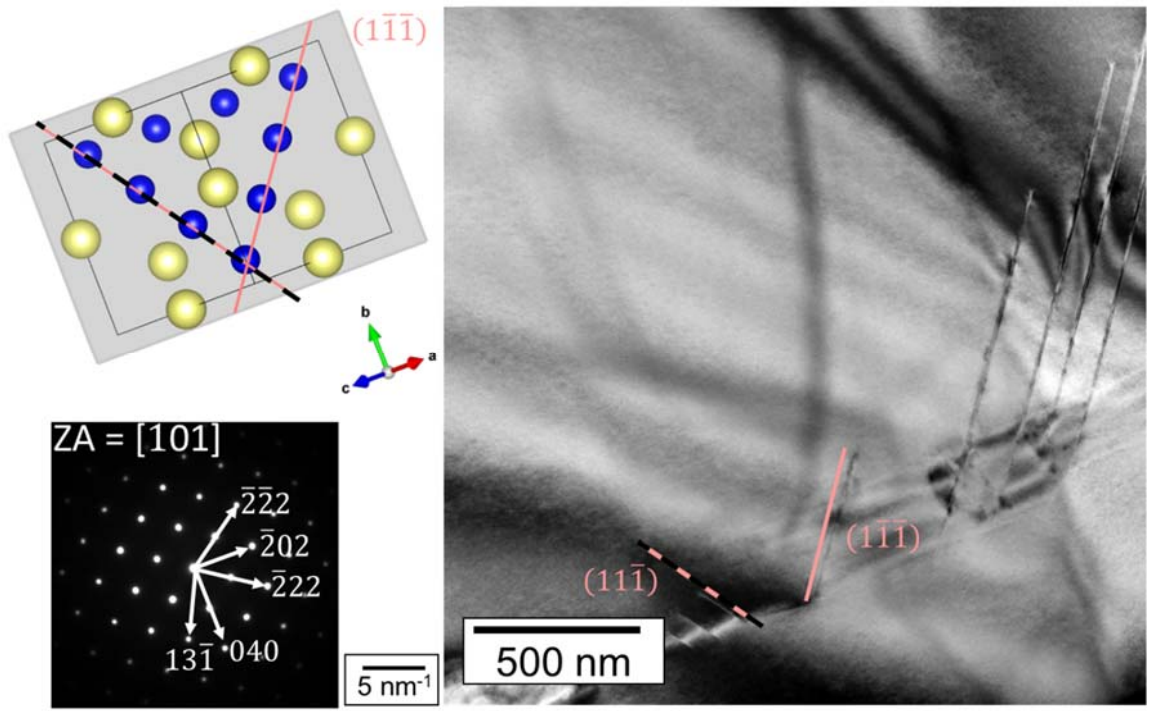

Figure 7: {111} stacking fault (SF) structures in a TEM BF image of the lower part of the Ca$_{36}$Al$_{53}$Mg$_{11}$ lamella in [101] zone axis, having two different orientated parallel assigned lines with a 70° rotation between them. In a) these lines are shown in the unit cell, being parallel to the $(11\bar{1})$ and $(1\bar{1}\bar{1})$ plane, visualised using VESTA [91].

## 3.5 Atomistic simulations

The effects of the chemical composition on the elastic properties of *C*15 Ca-Al-Mg Laves phases, including Young's modulus *E*, bulk modulus *B*, shear modulus *G*, and Poisson's ratio *v*, were investigated using atomistic simulations. For the stoichiometric *C*15 CaAl$_2$ phase, *B* = 52.5 GPa, *E* = 91.7 GPa, *G* = 37.9 GPa, and *v* = 0.209. As the Mg content increases to 6 at.% (corresponding to the chemical composition of Ca$_{33}$Al$_{61}$Mg$_6$), the values of *B*, *E*, and *G* decrease to 51.7, 88.1, and 36.2 GPa, respectively, as shown in Figure 8. By introduction of an over-stoichiometric Ca content, corresponding to the experimentally studied sample Ca$_{36}$Al$_{53}$Mg$_{11}$, the reduction of elastic moduli becomes more prominent, i.e., the values of *B, E,* and *G* decrease to 50.4, 81.3, and 33.0 GPa, respectively.

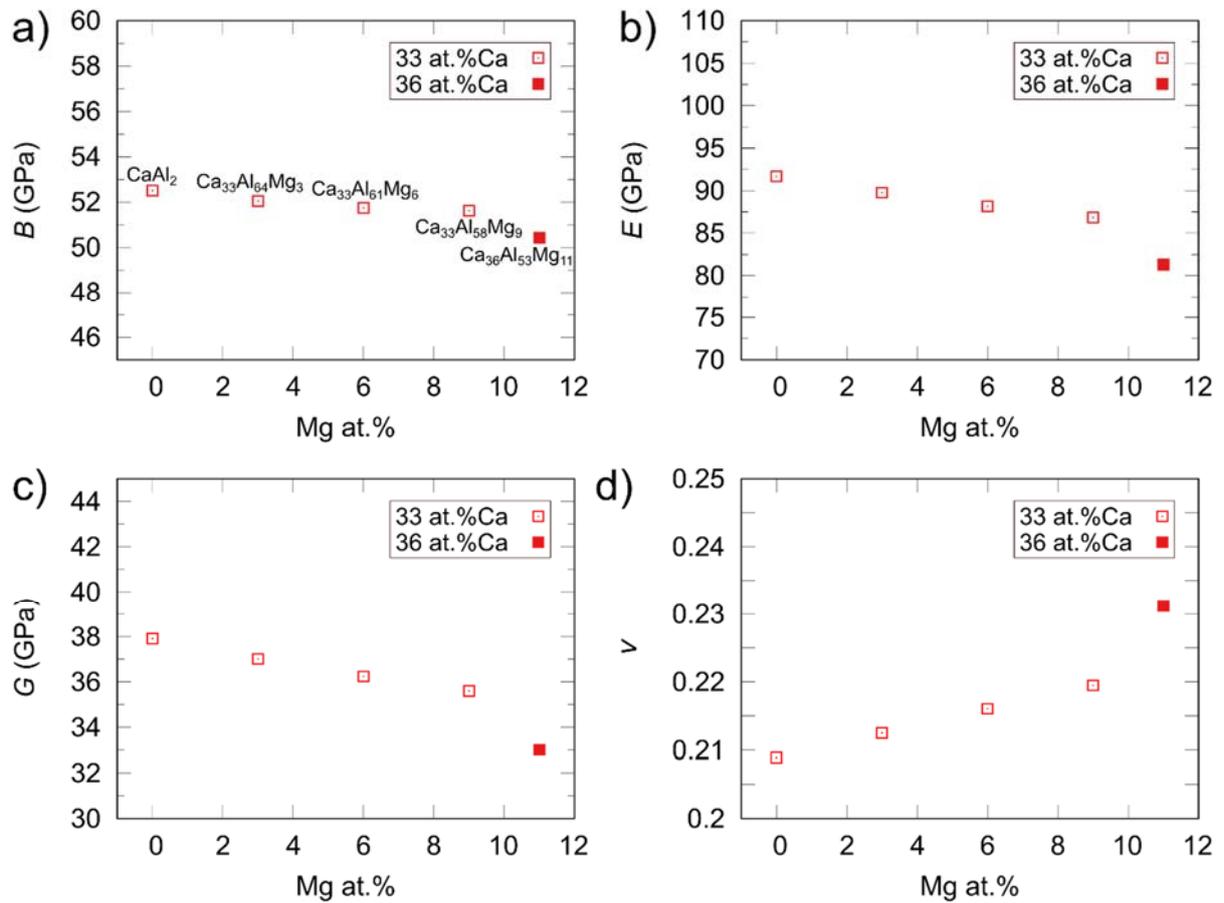

Figure 8: Elastic properties calculated through atomistic simulations on C15 Ca-Al-Mg Laves phases. a) Bulk modulus ($B$), b) Young's modulus ($E$), c) shear modulus ($G$) and d) Poisson's ratio ($v$).

The influence of the orientation-dependent behaviour of the mechanical properties is listed in Table 2 for the resulting Young's modulus and universal anisotropy index $A_U$ [92]. Values for the [111], [110] and [100] oriented Young's moduli for the tested compositions are given. Overall, the decreasing values can be seen for every composition, starting from [111] → [110] → [100], and from $Ca_{33}Al_{67}$ → $Ca_{33}Al_{61}Mg_6$ → $Ca_{36}Al_{53}Mg_{11}$. Additionally, with increasing Mg content the anisotropy index increases from 0.009 to 0.022.

Table 2: Calculated orientation dependent Young's modulus and anisotropy factor ($A_U$) for the three compositions:

|  | $Ca_{33}Al_{67}$ | $Ca_{33}Al_{61}Mg_6$ | $Ca_{36}Al_{53}Mg_{11}$ |
|---|---|---|---|
| $E_{[100]}$ | 87.9 GPa | 83.7 GPa | 76.0 GPa |
| $E_{[110]}$ | 92.6 GPa | 89.2 GPa | 82.6 GPa |
| $E_{[111]}$ | 94.3 GPa | 91.2 GPa | 85.0 GPa |
| $A_U$ | 0.009 | 0.013 | 0.022 |

In order to form a direct connection between the experimental and computational results, we further investigated the energy changes during many different potential slip events on the experimentally observed slip systems. To this end, we first compare all slip systems in $Ca_{33}Al_{67}$ before approaching the effect of chemical composition for a subset of these planes. The experimentally identified slip systems were assessed by calculating the corresponding GSFE lines (Figure 9a and γ-surfaces (Figure S2) as well as the correlated minimum energy paths via the NEB calculations (Figure 9b) in $C15$ $CaAl_2$. The energy barriers for all slip systems are summarized in Table S 2 in the supplementary material. For the GSFE calculation of the $1/6\,[\bar{2}1\bar{1}]$ partial slip on the $(11\bar{1})$ triple layer, the rigid-body shift was interpolated according to the path of the synchro-shear slip mechanism [39]. The energy barriers based on the GSFE lines for all slip systems range from approximately 1000 to 1600 $mJ/m^2$, where the $(001)\,[\bar{1}10]$ exhibits the highest barrier (1570 $mJ/m^2$) and $(11\bar{1})_{tk}[\bar{1}10]$ slip, that is slip between the triple and kagome layer (index $tk$) exhibits the lowest (1081 $mJ/m^2$) energy barrier among all slip events. The slip events along the minimum energy path (MEP) were obtained via the NEB calculation. The synchro-shear slip event ($(11\bar{1})_t[\bar{1}10]$, index $t$ to indicate synchro-shear taking place in the triple layer) exhibits the greatest change in energy level with the activation energy of this mechanism becoming the lowest (907 $mJ/m^2$) among all calculated slip events, as shown in Figure 9b and Table S1. Notably, $(001)\,[\bar{1}10]$ slip retains the highest activation energy (1564 $mJ/m^2$) along the MEP, followed by $(110)\,[\bar{1}10]$ slip (1210 $mJ/m^2$). The energy barriers along the MEP for other $(11\bar{n})\,[\bar{1}10]$ slip events range from 983 to 1190 $mJ/m^2$. Among all slip events considered here, stable stacking fault states exist only along the MEP of the synchro-shear slip event ($(11\bar{1})_t[\bar{1}10]$).

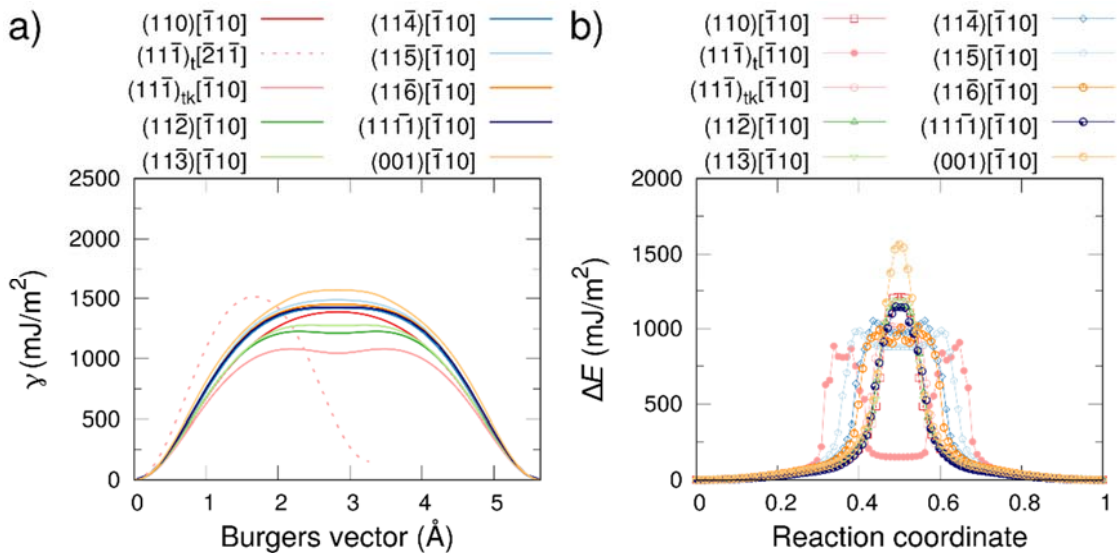

Figure 9: Assessment of $[\bar{1}10]$ slip systems in C15 $CaAl_2$ Laves phase. a) GSFE (γ) lines of full or partial $[\bar{1}10]$ slip on different slip planes. b) Excess energy ($\Delta E$) versus reaction coordinate of $[\bar{1}10]$ slip calculated using NEB.

The investigation into the effects of chemical composition on plasticity involved the calculation of GSFE lines for selected slip systems, varying solute concentrations and distributions in *C*15 Ca-Al-Mg Laves phases. Across all considered slip systems, the energy barriers decrease in off-stoichiometric compositions with an increase in Mg content and the lowest values are again obtained for the $Ca_{36}Al_{53}Mg_{11}$ composition (with over-stoichiometric Ca content), see Figure 10.

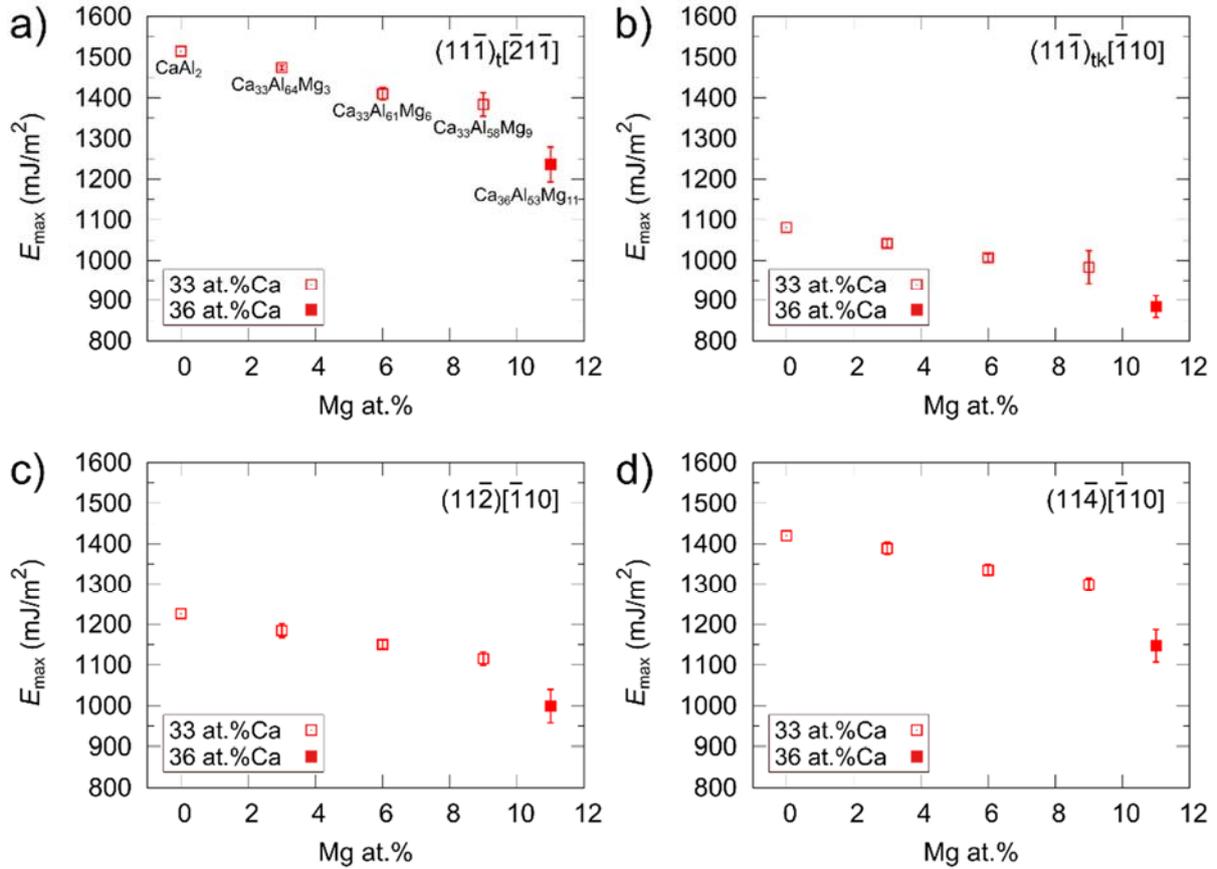

Figure 10 Energy barriers ($E_{max}$) in GSFE lines of $[\bar{1}10]$ slip in stoichiometric and off-stoichiometric C15 Ca-Al-Mg Laves phases. a) Partial $[\bar{1}10]$ (1/6 $[\bar{2}1\bar{1}]$) slip on the $(11\bar{1})$ triple layer along the synchro-shear slip path. b) Full $[\bar{1}10]$ slip on the $(11\bar{1})$ triple-Kagomé layer along the crystallographic slip path. Full $[\bar{1}10]$ slip on c) $(11\bar{2})$ and d) $(11\bar{4})$ planes.

Additionally, the synchro-shear-induced {111} stacking fault energy dramatically decreases by 50% from 123.2 mJ/m² in the stoichiometric composition to 60.4 ± 9.6 mJ/m² in the off-stoichiometric $Ca_{36}Al_{53}Mg_{11}$ composition as shown in Figure 11.

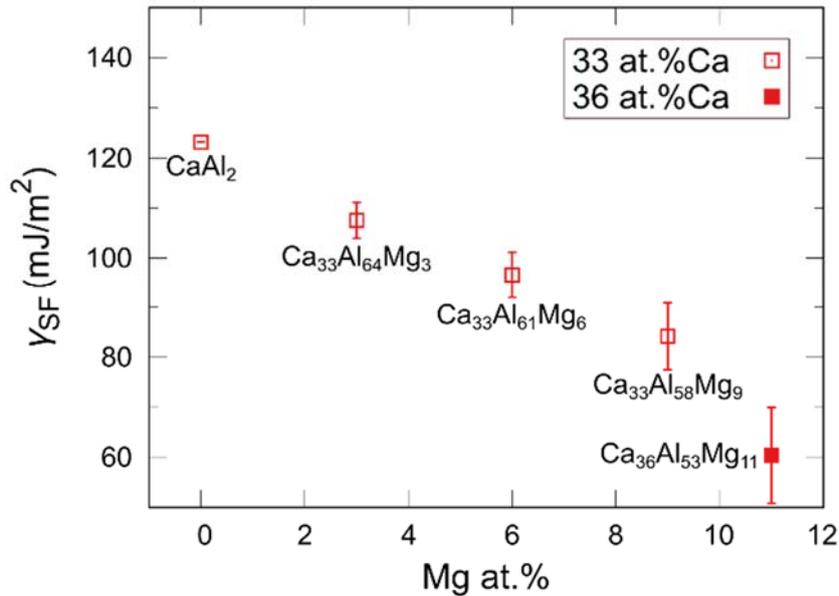

Figure 11 Synchro-shear-induced {111} stacking fault energies ($\gamma_{SF}$) in stoichiometric and off-stoichiometric C15 Ca-Al-Mg Laves phases.

## 4. Discussion

### 4.1 Phase analysis

We confirmed the presence of the cubic $C$15 Laves phase for both samples, $Ca_{33}Al_{61}Mg_6$ and $Ca_{36}Al_{53}Mg_{11}$, using both TEM and EBSD, with chemical compositions for these samples determined by EDX to 61.4 ± 0.2 at.% Al, 32.9 ± 0.2 at.% Ca and 5.7 ± 0.2 at.% Mg and 52.9 ± 0.7 at.% Al, 36.3 ± 0.1 at.% Ca and 10.8 ± 0.8 at.% Mg, respectively. Amerioun et al. [44] and Tehranchi et al. [93] used first principles calculations to estimate the stability range of $CaAl_{2-x}Mg_x$, i.e. for a stoichiometric Ca content. They proposed a stability limit for the $C$15 phase near 8 at.% and 13 at.% Mg, respectively at 0 K, that is either below or above in terms of the Mg content found here in the $C$15 phase, with the 13°at.% Mg both compounds fits. At 0 K Tehranchi et al. [93] conclude from their data that the $C$15 phase should directly decompose into a mixture of $C$15 + $C$14 phase as the $C$36 phase remains above the hull at 0 K. Hallstedt and Moori [94] note in their work on the Mg-Al-Ca ternary phase diagram that the energy difference between $C$15 and $C$36 is often small and the phase stability window therefore remains uncertain in particular for the $C$36 phase, which they also propose decomposes into $C$15 + $C$14 for temperatures below 200 °C. Zubair et al. [95] considered the three Laves polytypes as part of a metallic-intermetallic composite and found phase transitions between as-cast and heat-treated conditions. Most of their Laves phase precipitates proved under-stoichiometric in terms of Ca, however, the highest Ca-content of 34.2 at.% was found in a $C$15 precipitate (after heat treatment for 48h at 500 °C and subsequent slow cooling) with a 4.6 at.%

Mg and 61.2 at.% Al. These works on phase stability of the $C$15 Ca(Al,Mg)$_2$, together with our results, indicate that the Mg content investigated here are stable in the $C$15 phase, and that the slightly increased Ca content also appears to be contained within the same phase width. In the present work, we do not observe any indication of a bulk phase transformation to any of the hexagonal Laves phases., Although individual stacking faults on {111} planes were observed in the Ca$_{36}$Al$_{53}$Mg$_{11}$ and the stacking fault energy was found to drop with the addition of Mg and Ca in our atomistic simulations, consistent with expectations based on the first principles studies [44], no transition in the form of e.g. massive stacking fault formation could be observe. In terms of the mechanical properties, we therefore interpret all our data in the context of the $C$15 phase, but with the influence on the plastic deformation of the phase stability and the phase transformation path via synchro-shear in the triple layer in mind [16, 42, 43, 52].

## 4.2 Mechanical properties

We obtained indentation data for various orientations and three compositions when including the related binary C15 Ca$_{33}$Al$_{67}$ [24]. Average hardness and modulus decreased slight, albeit not statistically significantly through addition of Mg in Ca$_{33}$Al$_{61}$Mg$_6$ and a notable decrease was found for Ca$_{36}$Al$_{53}$Mg$_{11}$, with hardness and indentation modulus each approximately 16% lower than Ca$_{33}$Al$_{67}$.

Atomistic simulations corroborate the experimental findings and demonstrate that a substation of Al for Mg decreases hardness and elastic moduli values. The trend in calculated bulk elastic properties, follows a pattern with increasing Mg content (Figure 8), which is very consistent with the experimental findings.

The trend identified in this study aligns with the measurements of Luo et al. [35], who also found a softening behaviour, that is a decrease in hardness, with the deviation from the stoichiometric NbCo$_2$ $C$15 phase by increasing either Nb or Co content [35]. This softening behaviour can be interpreted by considering the effect of lattice distortion caused by introducing misfit solute atoms on plastic deformation. For the current phase, with the atomic radius of Mg being 17 pm larger than that of an Al atom, taken the atomic radius forming in the Laves phase [96], the incorporation of Mg atoms into the sublattice of Al would lead to change of Laves phase structure and may cause lattice distortion. The atomistic mechanisms of the softening effects by anti-site defects in the $C$14 CaMg$_2$ Laves phase have been demonstrated using atomistic simulations [41]. The activation energy of kink-pair nucleation, which serves as the rate-limiting step of synchro-Shockley dislocation motion on the basal or {111} plane, can be lowered by the presence of Ca$_{Mg}$ or Mg$_{Ca}$ anti-sites at the dislocation core region. This softening behaviour in off-stoichiometric compositions was also demonstrated for other {11n} slip systems (see

Figure 10), the mechanisms through which anti-site effects influence dislocation motion and the associated activation energies require further investigation.

Previous studies on the influence of ternary element additions have shown that the hardness and modulus typically remain stable or are minimally affected until surpassing a certain threshold of the additional element [35, 45, 47, 50, 51, 58, 72]. For instance, the addition of V to NbCr$_2$ at low concentrations (3 and 5 at.%) did not significantly alter mechanical properties, despite an increase in lattice parameter due to the replacement of smaller Cr atoms by V atoms [47, 58]. However, higher V concentrations (10, 18, and 25 at.%) elevated the brittle to ductile transition temperature (BDTT) [59], indicating that the amount of the added element is critical for property changes. This suggests that the promotion of dislocation mobility by solutes occurs only when their concentration exceeds a certain threshold, beyond which they lower the kink-pair nucleation barrier along a synchro-Shockley dislocation. Alternatively, if the added element initially occupies pre-existing structural vacancies, it may determine the barrier of property change. Moreover, increasing the atomic radius mismatch, as seen with the addition of Ca, can lead to a higher concentration of structural vacancies in Laves crystal structures, thereby facilitating synchro-Shockley dislocation motion compared to anti-site defects. These factors may explain why the Ca$_{36}$Al$_{53}$Mg$_{11}$ composition, with additional Ca and Mg, exhibits more significant changes in hardness and modulus compared to Ca$_{33}$Al$_{61}$Mg$_6$ with only Mg substitutions. Additionally, phase stability influences mechanical properties; for example, ab-initio calculations predict the formation of the ternary C36 phase of Ca(Al,Mg)$_2$ with a specific Mg content, suggesting a transition in properties [44].

## 4.3 Orientation dependence of mechanical properties

We also assessed orientation dependence in two ways: (1) by rotating the triangular impression of the indenter by 30° around the indentation axis and (2) by indenting along different crystal orientations. The first gave negligible changes in properties and no qualitative differences in the formation of slip traces or the distribution of identified planes apart from an overall greater number of slip traces. The assessment of orientation-dependent hardness and modulus was carried out in greatest depth on the intermediate sample in terms of composition, Ca$_{33}$Al$_{61}$Mg$_6$, which allowed the placement of many indents into each of the large grains. This is in contrast to the other samples, for example in case of the Ca$_{36}$Al$_{53}$Mg$_{11}$ only 4 to 6 indents per orientation could be placed. For Ca$_{33}$Al$_{61}$Mg$_6$, the hardness and modulus were consistent in 7 out of 8 areas, with only area III, closest to [111] (Figure 2) diverging towards higher values. This area gave a hardness of 5.5 ± 0.4 GPa and an indentation modulus of 89.5 ± 1.9 GPa. We assume that this may be due to two reasons. First, for the hardness, the absence of visible slip traces around the indent (Figure 2c), indicating the absence of easily activated slip planes for this orientation, which revealed only crack nucleation. A similar behaviour was reported for the stoichiometric Laves phase in a similar orientation [24], and it will be further discussed in the next section. Second, the influence of the anisotropy of this phase. Atomistic simulation calculates the modulus for all three compositions (Table 2) which shows that the Young's modulus reveals the highest values in [111] direction and the lowest at [100], at values of 91.2 GPa and 83.7 GPa, respectively, matching with the experimental data for $Ca_{33}Al_{61}Mg_6$. Additionally, increasingly anisotropic facture with increasing deviation from the binary composition was observed.

### 4.4 Deformation mechanisms

For the assessment of the activated slip planes, the overall activation frequencies were plotted for nine slip planes: the {100}, {110}, {111}, {112}, {113}, {114}, {115}, {116} and {1 1 11} planes. The alignment of these planes is displayed in Figure 3. All planes were further identified and confirmed with TEM investigations in Figure 6. In addition, the Burgers vectors identified by TEM ($\frac{1}{2}$ [101] and one of $\frac{1}{2}$ [110] or $\frac{1}{2}$ [01$\bar{1}$] on the (1 11 $\bar{1}$) and (16$\bar{1}$) planes as shown in Figure 3) were also confirmed by the atomistic simulations. The gamma surfaces and NEB calculations results indicate that no stable SF state exists as a result of simple shear or on the MEPs in {11n} planes, where n ≥ 2, revealing full dislocation slip as the dominant dislocation mechanism on these planes (while partial formation is possible on the {111} planes containing the triple layer).

For $Ca_{33}Al_{61}Mg_6$, two different indentation tests were performed (initial state and +30° rotated sample) in order to analyse the influence of the indenter geometry on the resulting slip activation. The rotation $Ca_{33}Al_{61}Mg_6$ only induces measurable differences in the amount of occurring slip planes, i.e., the second indentation (with the +30° rotated sample) results in 42% higher slip activation comparable to the first indentation (initial state). However, the type and distribution of planes remains the same.

Considering now the overall slip plane activation of $Ca_{33}Al_{61}Mg_6$, $Ca_{36}Al_{53}Mg_{11}$ and also $Ca_{33}Al_{67}$ [24], they all yield the dominant slip planes as the {11n} planes with the exception of the {111} plane. It is important to note that the activation frequencies stated here for $Ca_{33}Al_{67}$ deviate from those previously reported in [24], especially obvious for the reduced activation of {111} planes. This results from a change in the analysis protocol in that here we not only allowed double indexation in terms of different slip planes possessing a consistent surface trace orientation within the threshold (which remained constant), but we also counted individual planes of the same family separately. This allows us to not only weigh the relative occurrence of the different planes but also to assess their activation frequency with respect to the number of different available planes within a given family of planes, as indicated in the diagrams showing the activation frequencies of the different slip and crack planes (Figure 3b and Figure 3c). For example, for the {11n} plane family with n ≥ 2, the geometric probability of activation is higher, as these have 12 independent slip systems, compared to 4 for the {111} and 6 for the {100} and {110} plane families. A comparison of the relative activation frequencies in the view of this number of available slip planes gives a first indication of those planes which are easier or harder to activate than the average slip plane, as confirmed by a comparison of a slip trace analysis and micro compression in other intermetallics including the hexagonal $CaMg_2$ Laves phase [89, 98]. Here, we note that the relative activation of {110}, {111} and {11n} follows the number of available slip planes for all three samples with the sole exception the slightly higher activation of {111} slip in the binary $Ca_{33}Al_{67}$ sample, which lies in the error bar. Also consistent across all three samples is the reduced activation of slip and cracking on the {100} planes, as these lie lower than the {110} plane family that offers the same number of slip planes and below half the value measured for the {11n} planes with twice the number of independent slip planes.

Atomistic simulations confirm the experimental observation of the slip line statistics from the perspective of the expected energy barrier for slip. The minor detection of dislocation activity on the {001} planes is entirely consistent with the highest energy barrier on this plane both for simple shear and following the MEP as shown in Figure 9a and b. Slip on the {110} plane may be expected to occur in principle based on the atomistic simulations, as the energy barrier is the second highest along the MEP, but of the same order of magnitude as for the other {11n} planes. However, the plane was not observed in our TEM analysis. This finding is consistent with our slip trace analysis when considering the sampled orientations and slip trace morphology. Overall, the most dominantly activated slip planes are the {11n} planes with n ≥ 2 (Figure 3), showing slight differences that depend on the grain orientation. For the activation of the different slip planes differentiated by orientation, the reader is referred to Figure S1 in the supplementary materials.

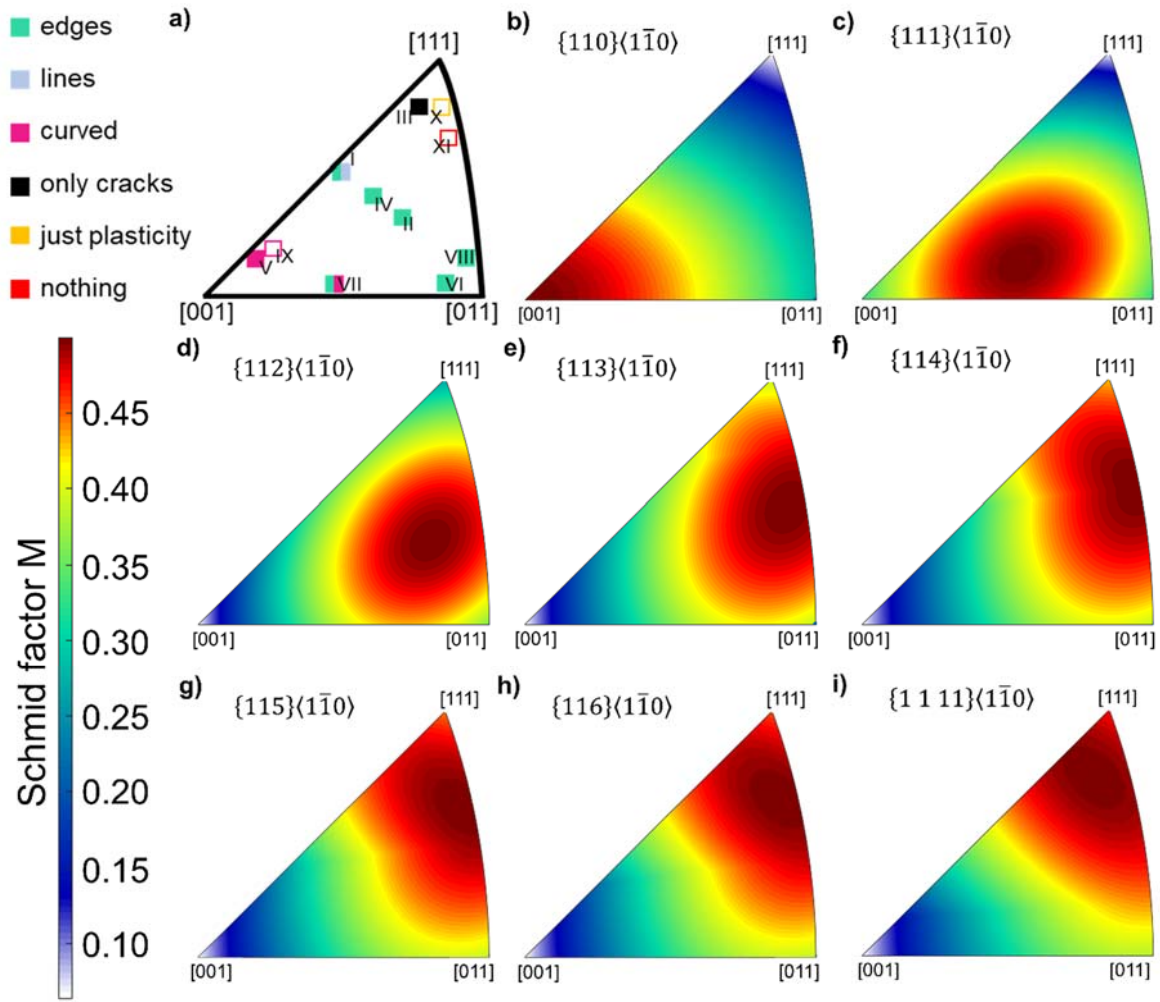

Figure 12: a) Orientation dependent slip morphology from Figure 3 with the Schmid factor M of the {110} – {1 1 11} slip systems (b)-i)) for deformation along the surface normal.

Figure 12 is broken down for the analysable orientations, the orientation-dependent Schmid factor M of the {11n} slip systems (Figure 12c-h) including the {111} plane (Figure 12b), is visualised in the IPF using a colour code for showing the magnitude of the calculated M. Here, the distribution of its maximum shows for the {11n} system starting with n = 2 and ending with n = 11, that the corner is shifting with increasing n from the location between [011] and [111], where the analysable surface traces (edges and straight lines) were located (Figure 12a), to [111]. For all these slip systems the [001] part always reveals the lowest M. For the $\{111\}\langle 1\bar{1}0\rangle$ system the maximum of M is central between [001] and [011], showing the lowest M at [111] but also a relatively low M for [001]. The highest M is present in <100> orientation. However, the slip trace morphology of indentations indented along a surface normal near [100] show curved slip traces, indicative of no single easily activated slip system. If {110} were viable slip planes with a comparable critical resolved shear stress and highest Schmid factor (assuming compression along the indentation axis) compared to the other {11n} planes, straight slip traces along {110} traces would be expected in this particular orientation, but are not found for any of the three compositions. We therefore assume that indexation of both {100} and {110} predominately results from the ambiguity of the method with different planes producing identical or closely aligned surface intersections depending on crystal orientation [77]. Taking a look at the deviation angle between the surface trace and the orientation of the plane trace, it is observable, that the counted {110} plane reveals a slightly different angle than one of the {11n} planes and lays often in the range of the {1 1 11}, {115}, {113} as well as the {112} plane. The other {11n} slip systems exhibit similar energy barriers, consistent with similar activation frequencies in nanoindentation tests for n ≥ 2 and a proportionally lower frequency for {111} with a third of the number of distinguishable planes for the two Ca(Al,Mg)$_2$ alloys.

So far, we have purposefully dealt with the {112} to {1 1 11} planes as one group of {11n} planes, as distinguishing these planes is difficult in experiments outside of edge-on viewing with correlated diffraction information in the TEM, which revealed the activation of these planes. In the slip trace analysis, an additional difficulty arises from the alignment of the plane traces and small deviations between these planes. For different n in the {11n} index, the angle between these planes continuously decreases with increasing n, i.e., the deviation between $(11\bar{1})$ and $(11\bar{2})$ is 20°, which decreases to 3° between $(11\bar{5})$ and $(11\bar{6})$. For the slip line analysis, the threshold for assignment of a possible slip plane to a slip trace was set to 3°, i.e. the same order of magnitude as the difference in inclination between the planes themselves (not necessarily the surface traces, which may lie at even more similar angles).

The difficulty of exact slip plane determination is further visualised in Figure 12, in that the distribution of the resolved shear stress becomes more similar as the planes' orientations converge in the higher order {11n} systems. The close alignment may of course result in a relative overestimation of the higher order {11n} planes where these are counted together perhaps more frequently than low index planes. However, we note that while these aspects result in difficulties in differentiation individual slip planes for large n, neither on its own is sufficient to result in indexation of these planes. If the planes possess substantially different critical resolved shear stresses, the similar Schmid factor would result in clear preferences for those with a lower activation barrier and the difference in alignment exceeds the experimental threshold with respect to the entire range of planes and in particular the lower order {11n} planes in most orientations.

We therefore conclude that the majority of plasticity is carried on the {11n} planes with n ≥ 1. Two further aspects remain to be considered: the effect of chemistry on the activation on these planes (and competing planes for fracture) and the likely slip system operating on the {111} planes, which is the only plane on which partials and the synchro-shear mechanism may operate.

As TEM confirms the presence of the new {11n} slip planes but indentation suggests very similar critical stresses for these planes, we refer to atomistic simulations in interpreting and bringing these results together to understand the dislocation-mediated plasticity in the Ca-Al-Mg Laves phases. The simulation results concerning the related energy barriers suggest that for the experimentally detected dominant {11n} plasticity, all barriers decrease with the addition of Mg and Ca (see Figure 10) in addition to the drop in elastic moduli (see Figure 8). Plastic deformation therefore becomes easier with increasing off-stoichiometric components and the stoichiometric $C$15 sample ($Ca_{33}Al_{67}$) exhibits the highest energy barriers for all slip events. This potentially explains the observed increasing dislocation and slip band density from the $Ca_{66}Al_{67}$ phase to both $Ca_{33}Al_{61}Mg_6$ and $Ca_{36}Al_{53}Mg_{11}$. In addition, the observation of dislocation dipoles is consistent with more easily activated plasticity the $C$15 phase containing additional Mg [99, 100], as dipoles may form more readily when the line energy is comparatively lower, allowing the more mobile dislocation structures to interact and rearrange. The computational results also reveal that by introducing Ca, the energy barriers significantly decrease and deviate from the linearly descending trend observed when only substituting Mg. This is entirely consistent with the sharper drop in hardness and modulus observed for the $Ca_{36}Al_{53}Mg_{11}$, which in terms of Mg addition increases by about the same increment as $Ca_{33}Al_{61}Mg_6$ relative to the binary alloy, but leads to a drop in hardness and stiffness that is more than twice as large. Furthermore, the finding that the energy barriers drop similarly on all systems is consistent with a chemistry-independent activation of the different slip planes in the statistical indentation slip trace analysis, particularly as they all share the same Burgers vector according to our TEM and atomistic simulation results, enabling cross-slip between them.

In close relation to these observations of {11n} plasticity and orientation dependence of flow, we also note that, for all three compositions, the indented orientation near the [111] direction exhibits no or very few surface slip traces. In $Ca_{33}Al_{61}Mg_6$ and $Ca_{33}Al_{67}$, cracks were found instead, but not in $Ca_{36}Al_{53}Mg_{11}$. In the [111] direction, the Schmid factors for the {11n} slip systems are high, with the maximum found for the {1 1 11} planes. The fact that no distinct surface plasticity is observed leads to the hypothesis that the resolved (normal) stress for crack nucleation on easy fracture planes is lower than that for dislocation motion. Therefore, the competition between cracking and slipping is resolved in favour of cracking. This is consistent with several planes lying nearly or, in case of the {112} planes directly, normal to the sample surface, giving a high elastic stress release during crack opening along the circumferential tensile component of the stress field during indentation.

Moreover, these [111]-like orientations exhibit higher values for hardness and indentation modulus as mentioned above (Table 1, area III and Table 2). The higher order {11n} planes lie closer together than lower index planes, giving overall not as many geometric options for slip on high Schmid factor planes in the three-dimensional stress field of the indentation along a

[111] axis, compared with the centre of the standard triangle, where the different lower index {11n} planes are also highly stressed. TEM and also the formation of curved surface traces provides additional indications that dislocations move under frequent cross-slip along the easiest planes around the imprinting indenter tip. In the {111} direction, the lack of variability in slip plane orientation coupled with the favourable orientation for cracking may therefore give rise to the observed transition towards cracking. Interestingly, $Ca_{36}Al_{53}Mg_{11}$ showed the same anisotropy in surface traces, but no cracks were observed in any orientation (Figure 2h). This is in contrast to $Ca_{33}Al_{61}Mg_6$ and $Ca_{33}Al_{67}$, where only cracks occurred in similar indentation orientations [24]. It could be rationalised by the decrease in the energy barrier for slip events (as illustrated in the atomistic simulations in Figure 10) in that the barrier for crack nucleation is not reached before plasticity takes place beneath the indent. In addition, the higher dislocation mobility may be associated with a lower rate of crack nucleus formation as dislocations are also found to occur on a greater number of slip planes, suggesting that lock formation and subsequent crack opening due to pile-up of dislocations is delayed in the $Ca_{36}Al_{53}Mg_{11}$ composition. Such an increase in the crack nucleation barrier with chemistry was also demonstrated by previous studies where the ternary Laves phase or the deviation from the stoichiometric composition results in higher fracture toughness [50, 54, 71, 101, 102], but not related as directly to changes in dislocation mobility and morphology.

Lastly, we consider the activation of synchro-shear in the cubic Laves phase. In the majority of the literature, this mechanism is assumed to constitute not just the major but sole dislocation mechanism accommodating plastic deformation in *C*15 Laves phases. Here, we find that slip among many different {11n} is vital to comprehending the cubic Laves phase's plasticity. Deformation on the {111} planes containing the triple layer is, however, also the only mechanism of dislocation motion that is associated with the occurrence of stacking faults or phase transformations between the different Laves polytypes.

Multiple previous studies on the dislocation structures of Laves phases with chemical variations reveal that the density of SFs is heavily influenced by the chemistry [35, 45-47, 72]. For the cubic *C*15 $NbCo_2$ Laves phase with a Nb content of 25.6 at.%, SFs occur on the {111} planes, whereas with increasing Nb content, the microstructure consists of dislocations and low-angle grain boundaries [35]. The same microstructural transition was also reported with the addition of ternary alloying elements to $NbCr_2$ by Yoshida et al. [45]. The structure remains a cubic Laves phase with the addition of relatively smaller atoms, while bigger atoms first reduce the stability of the cubic phase and then invoke micro-twins and SFs. In *C*15 Ca-Al-Mg Laves phases, the observation of SFs in $Ca_{36}Al_{53}Mg_{11}$ along the $(11\bar{1})$ and $(1\bar{1}\bar{1})$ planes, as shown in Figure 7, correlates well with the atomistic simulations, where a 50% decrease in $\gamma_{SF}$ was obtained in the off-stoichiometric sample with a similar composition as $Ca_{36}Al_{53}Mg_{11}$.

However, in spite of these observations of planar defect formation and dislocations aligned along {111} planes, it has been demonstrated that synchro-shear slip on the {111} plane is a thermally activated event in the sense that thermal assistance is indispensable for activating this slip event [40]. The required thermal fluctuation in atom positions is largely suppressed at room temperature in Laves phases with melting points much above 300 K, and hence, we tend to consider the {111} synchro-shear slip a high-temperature mechanism. This is the reason why overall the activation frequency of the {111} plane is lower than that of other {11n} planes despite having a similar energy barrier level.

## 5. Conclusions

The influence of chemical composition on plasticity in $C$15 Ca-Al-Mg Laves phases at room temperature was investigated using a combination of nanoindentation tests to obtain a statistical distribution of activated slip and crack planes, along with TEM analysis to characterize the introduced dislocation structures. Atomistic simulations were employed to reveal the energy barriers associated with these slip events. This work leads us to the following conclusions:

- The addition of Mg and Ca leads to a decrease in the hardness and indentation modulus relative to the binary composition.

- The cubic Ca(Al,Mg)$_2$ Laves phases are anisotropic with respect to stiffness and formation of surface traces from plastic deformation.

- The statistical analysis of the relative activation frequency of the surface traces revealed that the {11n} planes were the most activated slip planes, and this is rationalised by the comparably low energy barriers of these slip events obtained in atomistic simulations.

- {111} stacking faults were identified in the off-stoichiometric composition due to the significant decrease in stacking fault energy. However, it is not clear how this affects plastic properties as synchro-shear slip on the {111} plane requires thermal activation and, in agreement with this requirement, was found to be less activated compared to other {11n} planes across all compositions in our room temperature indentation experiments.

- Cracking was suppressed in nanoindentation of Ca$_{36}$Al$_{53}$Mg$_{11}$.

In summary, we find that the Ca(Al,Mg)$_2$ phase exhibits softening behaviour away from the CaAl$_2$ composition that is associated with distinct dislocation mechanisms, which we were able to infer using a combination of statistical indentation slip trace analysis, TEM and atomistic simulations. We believe that, in the future, this approach may also help elucidate the many conflicting results found elsewhere in the literature on chemistry-dependent Laves phase plasticity.

## 6. Acknowledgement

The authors gratefully acknowledge financial support by the Deutsche Forschunugsgemeinschaft (DFG) to all projects involved in this paper (A03, A05 and Z) of the SFB1394 Structural and Chemical atomic Complexity – From Defect Phase Diagrams to Material Properties, project ID 409476157. This project has received funding from the European Research Council (ERC) under the European Union's Horizon 2020 research and innovation programme (grant agreement No. 852096 FunBlocks). We would also like to express our gratitude to Marvin Poul and Jörg Neugebauer for early access to the Mg-Al-Ca machine learning potential used in this work.